\documentclass[11pt,draftcls,onecolumn]{IEEEtran}
\ifCLASSINFOpdf

\else

\fi

\usepackage{amssymb}  % assumes amsmath package installed
\usepackage{graphics} % for pdf, bitmapped graphics files
\usepackage{epsfig} % for postscript graphics files
\usepackage{mathptmx} % assumes new font selection scheme installed
\usepackage{times} % assumes new font selection scheme installed
\usepackage{amsmath} % assumes amsmath package installed
\usepackage{booktabs}
\usepackage{threeparttable}
\usepackage{floatrow}
\usepackage{multirow}
\usepackage{subfigure,url}
\usepackage{hyperref}
\usepackage{xcolor}
\newfloatcommand{capbtabbox}{table}[][\FBwidth]
\DeclareMathAlphabet{\mathcal}{OMS}{cmsy}{b}{n}
\DeclareMathAlphabet{\mathcal}{OMS}{cmsy}{m}{n}
\usepackage{boondox-cal}
\newcommand\numberthis{\addtocounter{equation}{1}\tag{\theequation}}
\newtheorem{theorem}{\indent Theorem}
\newtheorem{lemma}{\indent Lemma}

\newtheorem{proposition}{\indent Proposition}

\newtheorem{remark}{\indent Remark}
\newtheorem{problem}{\indent Problem}

\begin{document}

\title{A two-stage solution to quantum process tomography:  error analysis and optimal design}
\author{Shuixin Xiao,
	Yuanlong Wang, Jun Zhang,
Daoyi~Dong, Gary J. Mooney, Ian R. Petersen, and
Hidehiro~Yonezawa

\thanks{This research was supported by the Australian Research Council Future Fellowship Funding Scheme under Project FT220100656, Discovery Project Funding Scheme under Project DP210101938, the Centres of Excellence under Grant CE170100012, and the National Natural Science Foundation of China (62173229, 12288201). }
\thanks{Shuixin Xiao is with School of Engineering and Technology, University of New South Wales, Canberra ACT 2600, Australia, and University of Michigan--Shanghai Jiao Tong University Joint Institute, Shanghai Jiao Tong
	University, Shanghai 200240, China, and also with School of Engineering, The Australian National University, Canberra, ACT 2601, Australia (e-mail: shuixin.xiao@anu.edu.au). }
\thanks{Yuanlong Wang is with Key Laboratory of Systems and Control, Academy of Mathematics and Systems Science, Chinese Academy of Sciences, Beijing 100190, China (e-mail: wangyuanlong@amss.ac.cn).}
\thanks{Jun Zhang is with 
	University of Michigan--Shanghai Jiao Tong University Joint Institute, Shanghai Jiao Tong
	University, Shanghai 200240, China (e-mail: zhangjun12@sjtu.edu.cn).}
\thanks{Daoyi Dong is with CIICADA Lab, School of Engineering, Australian National University, ACT 2601, Australia, and  School of Physics, The University of Melbourne, Parkville, Victoria 3010, Australia (e-mail: daoyi.dong@anu.edu.au).}
\thanks{Gary J. Mooney is with School of Physics, University of Melbourne, Parkville, Victoria 3010, Australia (e-mail: mooney.g@unimelb.edu.au).}
\thanks{Ian R. Petersen is with School of Engineering, Australian National University, ACT 2601, Australia (e-mail: i.r.petersen@gmail.com).}
\thanks{ Hidehiro Yonezawa is with RIKEN Center for Quantum Computing, Japan, and
	School of Engineering and Technology, University of New South Wales, Canberra ACT 2600, Australia, and also with Centre for Quantum Computation and Communication Technology, Australian Research Council, Canberra, ACT 2600, Australia (e-mail: hidehiro.yonezawa@riken.jp).}
}

%\markboth{Submitted to IEEE Transactions on Information Theory}%
\maketitle

\begin{abstract}  
	Quantum process tomography is
	a critical task for characterizing the dynamics of quantum systems and achieving precise
	quantum control. In this paper, we propose a two-stage solution  for both trace-preserving and non-trace-preserving quantum process tomography. Utilizing a tensor structure, our algorithm exhibits a computational complexity of $O(MLd^2)$   where $d$ is the dimension of the quantum system and $ M $, $ L $ represent the numbers of different input states and measurement operators, respectively. We establish an analytical error upper bound and then design the optimal input  states and the optimal measurement operators, which are both based on minimizing the error upper bound and maximizing the robustness characterized by the condition number. Numerical examples and testing on IBM quantum devices are presented to demonstrate the performance and efficiency of our algorithm.
	
\end{abstract}
%\endNoHyper

\IEEEpeerreviewmaketitle

\section{Introduction}
In the past decades, quantum science and technologies have significantly advanced  in many fields such as quantum computation \cite{DiVincenzo255,qci}, quantum communication \cite{Gisin2007,5550402}, quantum sensing \cite{RevModPhys.89.035002} and quantum control \cite{dong2010quantum,dong2022quantum,Dong2023}. To fully develop and realize these technologies, a fundamental challenge to overcome is to characterize unknown quantum dynamics. A typical framework to formulate this problem is quantum process tomography (QPT), where the parameters determining the map from input quantum states to output states need to be estimated.

For a closed quantum system, QPT is reduced to Hamiltonian identification and many results have been obtained in this scenario \cite{PhysRevLett.108.080502,PhysRevA.95.022335,9026783,PhysRevLett.113.080401,8022944,unitary}.
For an open quantum system characterized by time-independent parameters, the process  is completely positive (CP) and attention is usually restricted to the trace-preserving (TP) case.
Using direct and convex optimization methods, Zorzi \emph{et al.} \cite{Zorzi2014,6426789} gave the minimum positive operator-valued measure (POVM) resources required to estimate a CPTP quantum process,  and Knee \emph{et al.} \cite{cptp2018} and Surawy-Stepney \emph{et al.} \cite{surawy2021projected} proposed iterative projection algorithms to identify such a process. The efficiency of parameter estimation for quantum processes was studied in \cite{4655455}. Zhang and Sarovar \cite{zhang2} utilized the time traces of observable measurements  to  identify the parameters in a master equation. The identifiability and identification for the passive quantum systems were investigated in \cite{7130587}. When the quantum process is non-trace-preserving (non-TP), Huang \emph{et al.}  \cite{HUANG2020286} proposed a convex optimization method and Bongioanni \emph{et al.} \cite{PhysRevA.82.042307} demonstrated non-TP QPT based on a maximum likelihood approach in an experiment. In practice, a quantum process may be affected by non-Markovian  noise.  White \emph{et al.} \cite{white2020demonstration} developed a framework to characterize non-Markovian dynamics and the work in \cite{PRXQuantum.3.020344} proposed  process tensor tomography for non-Markovian QPT.
Moreover, adaptive strategies for QPT were proposed in \cite{Wang2016,inada,PhysRevA.95.012302}.

In this paper, we focus on Standard Quantum Process Tomography (SQPT) \cite{qci,unitary,jmo,ccqp,PhysRevA.63.020101} where states with the same dimension as the process are inputted, and the corresponding outputs are measured to reconstruct the process.
We extend the framework in \cite{qci,8022944} to allow for general input states and non-TP processes, and also directly connect the process with measurement data in a unified system equation. Using a mathematical representation, we deduce the special structural properties of the system equation, allowing for more efficient analysis, solution and optimal design.
We give an analytical two-stage solution (TSS) to a general QPT problem, as a closed-form estimation which is not common among the existing QPT solutions.  For $ d $-dimensional quantum systems, our algorithm has computational complexity $ O(MLd^2) $ and storage requirement $ O(ML) $, where $ M \geq d^2 $ is the number of different informationally complete or overcomplete input states and 
$ L\geq d^2$ is the number of different  measurement operators.
We also establish an analytical error upper bound for our algorithm. Then we study the optimization of input states and measurement operators, based on this error upper bound and the condition number of the system equation, which reflects the robustness of our algorithm.  We give theorems to characterize lower bounds on the error and condition number for both input states and measurement operators.
We prove that SIC (symmetric informationally complete) states and MUB (mutually unbiased bases) states \cite{xiao2021optimal} are two examples achieving the lower bounds for our algorithm.
We also prove MUB  measurement achieves the lower bounds among measurement operators.
Numerical examples and testing on IBM Quantum devices demonstrate the
effectiveness of our algorithm and validate the theoretical error analysis. 
We compare our TSS algorithm with the convex optimization method in \cite{unitary} and the results show that our algorithm is more efficient in both time and space costs during calculation. The main contributions
of this paper are summarized as follows.
\begin{enumerate}
	\item[(i)] We propose an analytical two-stage solution (TSS) to a general QPT problem. Our TSS algorithm  does not require specific prior knowledge about the unknown process and can be applied in both TP and non-TP quantum processes.
	\item[(ii)] Utilizing a tensor structure, our TSS algorithm  has the computational complexity $O(MLd^2)$ and  a clear storage requirement $O(ML)$, resulting from our closed-form estimation formula.
	\item[(iii)] Our TSS algorithm  can give an  analytical error upper bound which can be further utilized to 
    optimize the input states and the measurement operators.
	\item[(iv)] Numerical examples and testing on IBM Quantum devices demonstrate the theoretical results and the effectiveness of our method. Compared to the convex optimization method in \cite{unitary},  our algorithm is more efficient in both time and space costs.
\end{enumerate}

The organization of this paper is as follows. Section
\ref{Sec2} introduces some preliminary knowledge and establishes our framework for QPT. Section \ref{sec3} studies the tensor structure and presents our TSS algorithm for QPT. In Section \ref{sec4}, we analyze the computational complexity and the storage requirement. In Section \ref{sec5}, we show the error analysis and in Section \ref{optimal}, we study the optimization of the input  states and the measurement operators.
Numerical examples are presented in Section \ref{numerical} and Section \ref{sec8} concludes this paper. 

Notation:  The $ i $-th row and $ j $-th column of a matrix $ X $ is $X_{ij} $. The $ j $-th column of $ X $ is $ \operatorname{col}_{j}(X) $. The transpose of $X$ is $X^T$. The conjugate $ (*) $ and transpose of $X$ is $X^\dagger$. The sets of  real and complex numbers are $\mathbb{R}$ and $\mathbb{C}$, respectively. The sets of  $d$-dimension complex vectors and  $d\times d$ complex matrices are $\mathbb{C}^d$ and $\mathbb{C}^{d\times d}$, respectively. The identity matrix is $ I $. $\rm i=\sqrt{-1}$. 
The trace of $X$ is $\text{Tr}(X)$. The Frobenius norm of a matrix $X$ is denoted as $||X||$ and the 2-norm of a vector $ x $ is $||x||$.  We use density matrix $ \rho $ to  represent a quantum state where $ \rho=\rho^{\dagger} $, $ \rho\geq0 $ and $ \operatorname{Tr}(\rho)=1 $ and use a unit complex vector  $|\psi\rangle$ to represent a pure state. The estimate of $X$ is $ \hat X $. The inner product of two matrices $X$ and $Y$ is defined as $\langle X, Y\rangle\triangleq\text{Tr}(X^\dagger Y)$.
The inner product of two vectors $x$ and $y$ is defined as $\langle x,
y\rangle\triangleq x^\dagger y$. The tensor product of $A$ and $B$ is denoted $A\otimes B$. Hilbert space is $\mathbb{H}$. The Kronecker delta function is $\delta$. The $\operatorname{diag}(a)$ denotes a diagonal matrix with the $i$-th diagonal element being the $i$-th element of the vector $a$.  For any positive semidefinite $X_{d \times d}$ with spectral decomposition $X=U P U^{\dagger}$, we define $\sqrt{X}$ or $X^{\frac{1}{2}}$ as $U \operatorname{diag}\left(\sqrt{P_{11}}, \sqrt{P_{22}}, \cdots, \sqrt{P_{d d}}\right) U^{\dagger}$.

\section{Preliminaries and quantum process tomography}\label{Sec2}
\subsection{Vectorization function}
We introduce the (column-)vectorization function $\text{vec}: \mathbb{C}^{m\times n}\mapsto \mathbb{C}^{mn}$. For a matrix $X_{m\times n}$, 
\begin{equation}
\begin{aligned}
\text{vec}(X_{m\times n})\triangleq&[X_{11},X_{21},\cdots,X_{m1},X_{12},\cdots,X_{m2},\\
&\cdots,X_{1n},\cdots,X_{mn}]^T.
\end{aligned}
\end{equation}
Thus,  $ \text{vec}(\cdot) $ is a linear and basis dependent map.  We also define the inverse $\text{vec}^{-1}(\cdot)$ which maps a $d^2\times 1$ vector into a $d\times d$ square matrix.
The common properties of $ \text{vec}(\cdot) $ are listed as follows \cite{horn_johnson_2012,watrous2018theory}:
\begin{equation}\label{property1}
\langle X, Y\rangle=\langle \text{vec}(X), \text{vec}(Y)\rangle,
\end{equation}
\begin{equation}\label{property2}
\text{vec}(XYZ)=(Z^T\otimes X)\text{vec}(Y),
\end{equation}
\begin{equation}\label{property4}
\text{Tr}_1(\text{vec}(X)\text{vec}(Y)^{\dagger})=XY^{\dagger},
\end{equation}
where $\text{Tr}_1(X)$denotes the partial trace on the space $\mathbb H_1$ with $X$ belonging to the space $\mathbb H_1\otimes \mathbb H_2$.
\subsection{Quantum process tomography}
According to the system architecture, QPT generally can be divided into three classes: Standard Quantum Process Tomography (SQPT) \cite{qci,unitary,jmo,ccqp,PhysRevA.63.020101}, Ancilla-Assisted Process Tomography (AAPT) \cite{jmpcp,meaqo,aaqpt,ici} and Direct Characterization of Quantum Dynamics (DCQD) \cite{dcqd,dcgt,edcqy}. In SQPT, one inputs quantum states with the same dimension as the process, and then we reconstruct the process from quantum state tomography (QST) on the output states. In AAPT and DCQD, an auxiliary system (ancilla) is attached to the principal system, and the input states and the measurement of the outputs are both carried out on the extended Hilbert space. AAPT transforms QPT to QST in a larger Hilbert space and the input state can be separable \cite{qptre}.
DCQD requires entangled input states and measurements such that the measured probability distributions are directly related to the elements of the process representation \cite{qptre}. Since AAPT needs high dimensional states in the extended Hilbert space and DCQD  usually requires entanglement which are both more technologically demanding, we focus on the SQPT scenario in this paper.

Here, we  briefly introduce the SQPT framework as given in  \cite{qci,8022944} and extend it to a more general framework based on the following two aspects:
\begin{enumerate}
	\item[(i)] Input Fock states are extended to be arbitrary states.
	\item[(ii)] The unknown process is extended from the TP case to both the TP and non-TP cases.
\end{enumerate}

For a $ d $-dimensional quantum system, its dynamics can be described by a completely-positive (CP) linear map  $ \mathcal{E} $.  If we input a quantum state $\rho^{ \text{in}}$ (\emph{input state}), using the
Kraus operator-sum representation \cite{qci}, the output state $\rho^{ \text{out}}$ is given by 
\begin{equation}\label{Kraus}
\rho^{ \text{out }}=\mathcal{E}\left(\rho^{ \text{in }}\right)=\sum_{i=1}^{d^2} \mathcal A_{i} \rho^{\text{in }}\mathcal  A_{i}^{\dagger},
\end{equation}
where  $\mathcal A_{i} \in \mathbb{C}^{d \times d}$ and satisfy
\begin{equation}\label{aleq}
\sum_{i=1}^{d^2}\mathcal  A_{i}^{\dagger}\mathcal  A_{i}\leq I_{d}.
\end{equation}
When the equality in \eqref{aleq} holds, the map $ \mathcal{E}  $ is said to be trace-preserving (TP). Otherwise, it is non-trace-preserving (non-TP).
Let $\{E_i\}_{i=1}^{d^2}$ be a fixed basis of $ \mathbb{C}^{d \times d}$. 
From $ \{\mathcal A_{i}\}_{i=1}^{d^2} $, we can obtain  \emph{process matrix} $ X $ \cite{qci,8022944} which is a $ d^2\times d^2 $ Hermitian and positive semidefinite matrix such that
\begin{equation}
\mathcal{E}(\rho^\text{in})=\sum_{j,k=1}^{d^2}E_j\rho^\text{in} E_k^\dagger X_{jk}.
\end{equation}
In this paper, we consider both TP and non-TP quantum processes and
\eqref{aleq}  becomes
\begin{equation}\label{ptrace1}
\sum_{j,k=1}^{d^2}X_{jk}E_k^{\dagger}E_j\leq I_{d}.
\end{equation}
Since the relationship between $ X $ and $ \mathcal{E} $ is one-to-one \cite{qci,8022944},  the identification of  $ \mathcal{E}  $ is equivalent to identifying $ X $.
The number of independent elements in $ X $ is $ d^4-d^2 $ for the TP case and $ d^4 $ for the non-TP case.

Let $ \{\sigma_{n}\}_{n=1}^{d^2} $ be a complete basis set of $ \mathbb{C}^{d \times d} $. Denote the set of all the input states as $ \{\rho_m^{\text{in}}\}_{m=1}^{M} $.
We can expand each output state uniquely in $\{\sigma_{n}\}_{n=1}^{d^2} $ as
\begin{equation}\label{eqlambda}
\rho_{m}^\text{out}=\mathcal{E}(\rho_m^{\text{in}})=\sum_{n=1}^{d^2} \alpha_{mn}\sigma_{n}.
\end{equation}
We also define $ \beta_{mn}^{jk} $ such that
\begin{equation}\label{betadef}
E_j \rho_m^{\text{in}} E_k^\dagger=\sum_{n=1}^{d^2} \beta_{mn}^{jk}\sigma_n.
\end{equation}
From the linear independence of $\{\sigma_n\}_{n=1}^{d^2}$, the relationship between $X$ and $\alpha$ is \cite{qci,8022944}
\begin{equation}\label{one}
\sum_{j,k=1}^{d^2}\beta_{mn}^{jk} X_{jk}=\alpha_{mn}.
\end{equation}

To guarantee that \eqref{one} has a unique solution, the maximal linear independent subset of $ \{\rho_m^{\text{in}}\}_{m=1}^{M} $ must have $ d^2 $ elements. If this is achieved with $ M=d^2 $, we call the input state set \emph{informationally complete}. If this is achieved with $ M> d^2 $, we call it \emph{informationally overcomplete}. The work in \cite{qci,8022944} is restricted to a special complete case, where $\{\rho_m^{\text{in}}\}_{m=1}^{M}=\{|j\rangle\langle k|\}_{j,k=1}^{d}  $ (after a linear combination of the practical experiment results).
Here we extend the framework of \cite{qci,8022944} to the general informationally complete/overcomplete case with $ M\geq d^2 $.
We define the matrix $A$ where $A_{mn}\triangleq\alpha_{mn}$ and arrange the elements $\beta_{mn}^{jk}$ into a matrix $ B$:
\begin{equation}\label{matrixB}
B=\left[\begin{array}{*{7}{c}}
\beta_{11}^{11} & \beta_{11}^{21} & \cdots & \beta_{11}^{12} & \beta_{11}^{22} & \cdots & \beta_{11}^{d^2d^2} \\
\beta_{21}^{11} & \beta_{21}^{21} & \cdots & \beta_{21}^{12} & \beta_{21}^{22} & \cdots & \beta_{21}^{d^2d^2} \\
\multicolumn{7}{c}{\dotfill} \\
\beta_{12}^{11} & \beta_{12}^{21} & \cdots & \beta_{12}^{12} & \beta_{12}^{22} & \cdots & \beta_{12}^{d^2d^2} \\
\beta_{22}^{11} & \beta_{22}^{21} & \cdots & \beta_{22}^{12} & \beta_{22}^{22} & \cdots & \beta_{22}^{d^2d^2} \\
\multicolumn{7}{c}{\dotfill} \\
\beta_{Md^2}^{11} & \beta_{Md^2}^{21} & \cdots & \beta_{Md^2}^{12} & \beta_{Md^2}^{22} & \cdots & \beta_{Md^2}^{d^2d^2} \\
\end{array}\right]
\end{equation}
which is an $Md^2\times d^4$ matrix. We rewrite \eqref{one} into a compact form as
\begin{equation}\label{eq3}
B\text{vec}(X)=\text{vec}(A),
\end{equation}
where $  B$ is determined once the bases $\{E_i\}_{i=1}^{d^2}$ and $ \{\sigma_{n}\}_{n=1}^{d^2} $ are chosen.

In an experiment, assume that the number of copies for each input state is $ N $.  Assume $ P_{i,j} $ is the $ i $-th POVM element (i.e.,  measurement operator) in the $ j $-th POVM set where $1\leq j\leq J  $ and $ 1\leq i \leq n_j  $.  Therefore, the total number of different measurement operators is $ L=\sum_{j=1}^{J} n_{j} $  and we have
\begin{equation}\label{povmcom}
P_{1,j}+P_{2,j}+\cdots+P_{n_{j},j}=I_{d},
\end{equation}
for $ 1\leq j\leq J $, which is the completeness constraint for  each POVM set. 
We then vectorize these POVM elements  as $ \{P_l\}_{l=1}^{L} $ where, e.g., $P_{i+\sum_{q=1}^{j-1}n_q}\leftarrow P_{i,j}$. These are
the measurement operators on the output states. To extract all of the information in the output state, these measurement operators should be informationally complete or  overcomplete, and thus $ L\geq d^2 $. Some efficient methods for QST with fewer measurements are also discussed in \cite{6636034,7956181,6942238}.
Then, we can also uniquely expand  $ P_{l} $ in $ \{\sigma_{n}\}_{n=1}^{d^2} $ as
\begin{equation}
P_{l}=\sum_{j=1}^{d^{2}} \mu_{l j} \sigma_{j},
\end{equation} 
and the ideal measurement probability $p_{m l}$  of the $ m $-th output state and the $ l $-th measurement operator is
\begin{align*}
p_{m l}&=\operatorname{Tr}\left(\rho^\text{out}_{m} P_{l}\right)\\
&=\operatorname{Tr}\left(\sum_{n=1}^{d^{2}} \alpha_{m n} \sigma_{n} \sum_{j=1}^{d^{2}} \mu_{l j} \sigma_{j}\right)\\
&=\sum_{n,j=1}^{d^{2}} \alpha_{m n} \mu_{l j} \operatorname{Tr}\left(\sigma_{n} \sigma_{j}\right). \numberthis
\end{align*}
Define the matrix $ \mathcal{P} $ with $ \mathcal{P}_{ml}\triangleq p_{ml} $ and let
\begin{equation}
C\!\triangleq\!\!\left[\begin{array}{cccc}
\operatorname{Tr}\left(\sigma_{1} \sum_{j=1}^{d^{2}} \mu_{1 j} \sigma_{j}\right) & \cdots & \operatorname{Tr}\left(\sigma_{d^{2}} \sum_{j=1}^{d^{2}} \mu_{1 j} \sigma_{j}\right) \\
\operatorname{Tr}\left(\sigma_{1} \sum_{j=1}^{d^{2}} \mu_{2 j} \sigma_{j}\right)  & \cdots & \operatorname{Tr}\left(\sigma_{d^{2}} \sum_{j=1}^{d^{2}} \mu_{2 j} \sigma_{j}\right) \\
\vdots & \vdots & \vdots \\
\operatorname{Tr}\left(\sigma_{1} \sum_{j=1}^{d^{2}} \mu_{L j} \sigma_{j}\right)  & \cdots & \operatorname{Tr}\left(\sigma_{d^{2}} \sum_{j=1}^{d^{2}} \mu_{L j} \sigma_{j}\right)
\end{array}\right],
\end{equation}
which is determined by the measurement operators  $ \{P_l\}_{l=1}^{L} $.
Thus we have 
\begin{equation}\label{maineq2}
\left(I_{M} \otimes C\right) \operatorname{vec}(A^{T})=\operatorname{vec}(\mathcal{P}^{T}).
\end{equation}
Define $ K $ such that $ K\operatorname{vec}(A)=\operatorname{vec}(A^{T}) $ and thus $ K $ is a $ Md^2\times Md^2 $ commutation matrix.
Using \eqref{eq3} and  \eqref{maineq2}, we have
\begin{equation}\label{tensora}
\left(I_{M} \otimes C\right)K B\operatorname{vec}(X)=\operatorname{vec}(\mathcal{P}^{T}),
\end{equation}
which directly connects the unknown quantum process $ X $ with measurement data $ \mathcal{P} $ in the experiment.

Assume that the practical measurement result is $\hat p_{m l}$ and the measurement error is $ e_{ml}=\hat p_{m l}- p_{m l} $.
According to the central
limit theorem, $ e_{ml} $ converges in distribution to a normal
distribution with mean zero and variance $\left(p_{ml}-p_{ml}^{2}\right) /(N / J)$ \cite{Qi2013,MU2020108837}. We consider the general scenario where there is no prior knowledge about whether the process is TP or non-TP, and the QPT  can be formulated as the following optimization problem:
\begin{problem}\label{problem1}
	Given the parameterization matrix $B$ for the input states, the parameterization matrix $C$ for the measurement operators, the permutation matrix $ K $ and the experimental data $\hat{ \mathcal{P}}$, find a Hermitian and positive semidefinite estimate $\hat{X}$ minimizing
	$\left\|\left(I_{M} \otimes C\right) KB\operatorname{vec}(\hat{X})-\operatorname{vec}(\widehat{ \mathcal{P}}^{T})\right\|$
	and satisfying the  constraint \eqref{ptrace1}.
\end{problem}

\section{Tensor structure and two-stage quantum process tomography solution}\label{sec3}
In this section, we discuss how to simplify the structure of $ B $ under a suitably chosen representation, which can reduce the computational complexity and storage requirement. Then we propose a two-stage solution (TSS) to obtain analytical tomography formulas for Problem \ref{problem1}. When there is prior knowledge that the process is TP, most of the following analysis still applies, unless otherwise specified.
\subsection{Tensor structure of QPT}
Firstly, we consider the basis sets  $\{E_i\}_{i=1}^{d^2}$  and $ \{\sigma_{n}\}_{n=1}^{d^2} $, which, if properly chosen, can simplify the QPT problem and greatly benefit the time and space costs of our algorithm. We  choose these basis sets as the \emph{natural basis} $\{|j\rangle\langle k|\}_{1\leq j,k\leq d}$ where $i=(j-1)d+k, n=(k-1)d+j$. The advantages of this choice can be demonstrated as follows.
\begin{lemma}\label{ptracepropo}
	\cite{8022944,Zorzi2014}
	If $\{E_i\}_{i=1}^{d^2}$ is chosen as the natural basis  $\{|j\rangle\langle k|\}_{1\leq j,k\leq d}$ where $ i=(j-1)d+k $, then the constraint \eqref{ptrace1} becomes $\operatorname{Tr}_1(X)\leq I_d$.
\end{lemma}
Lemma \ref{ptracepropo} fellows from the Choi-Jamiołkowski isomorphism and the proof for $\operatorname{Tr}_1(X)= I_d  $ in the TP case can be found in \cite{8022944,Zorzi2014}, which can be straightforwardly extended to non-TP cases. Define $F \triangleq  \operatorname{Tr}_{1} (X)\leq I_{d}$, and  $F$  represents the success probability of the quantum process \cite{PhysRevA.82.042307}. Furthermore, we have the following benefit.
\begin{proposition}\label{btensor}
	Define the collection  of all the vectorized input states as
	\begin{equation}\label{probestate}
	V\triangleq\left[\operatorname{vec}\left(\rho_1^{\text{in}}\right), \operatorname{vec}\left(\rho_2^{\text{in}}\right), \cdots, \operatorname{vec}\left(\rho_M^{\text{in}}\right)\right].
	\end{equation}
	If $\{E_i\}_{i=1}^{d^2}$ and $ \{\sigma_{n}\}_{n=1}^{d^2} $ are chosen as the natural basis $\{|j\rangle\langle k|\}_{1\leq j,k\leq d}$ where $ i=(j-1)d+k, n=(k-1)d+j $,
	then we have
	\begin{equation}\label{structureb}
	\left(I_{d^{2}} \otimes V^{T}\right) R=B,
	\end{equation}
	where $ R $ is a $ d^4\times d^4 $ permutation matrix.
\end{proposition}
\begin{IEEEproof}
	Using \eqref{property2}, \eqref{betadef} becomes
	\begin{equation}
	\left(E_{k}^{*} \otimes E_{j}\right) \operatorname{vec}\left(\rho_m^{\text{in}}\right)=\sum_{n=1}^{d^2} \beta_{{mn}}^{{jk}} \operatorname{vec}\left(\sigma_{n}\right).
	\end{equation}
	Since  $ \{\sigma_{n}\}_{n=1}^{d^2} $ the is natural basis, using \eqref{probestate}, we have
	\begin{equation}
	\left(E_{k}^{*} \otimes E_{j}\right) V=\left[\begin{array}{ccc}
	\beta_{1 {1}}^{{jk}} & \cdots & \beta_{{M1}}^{{jk}} \\
	\vdots & \vdots & \vdots \\
	\beta_{1 {d^2}}^{{jk}} & \cdots & \beta_{{Md^2}}^{{jk}}
	\end{array}\right],
	\end{equation}
	and  $ \operatorname{vec}\left(\left[\left(E_{k}^{*} \otimes E_{j}\right) V\right]^{T}\right) $ is the $ (j+(k-1)d^2) $-th column of $ B $. Let $k=(u-1) d+v$, $j=(x-1) d+y$. Then
	\begin{equation}
	E_{k}^{*} \otimes E_{j}=|u\rangle\langle v|\otimes| x\rangle\langle y|,
	\end{equation}
	where only the element at $ \left((u-1)d+x, (v-1)d+y\right) $ is non-zero. Then
	\begin{equation}
	\begin{aligned}
	\left(E_{k}^{*} \otimes E_{j}\right) \operatorname{vec}\left(\rho_m^{\text{in}}\right)&=[0, \cdots, 0,\underbrace{\left(\rho_m^{\text{in}}\right)_{y v}}_{((u-1)d+x)\text{th}}, 0, \cdots, 0]^{T},
	\end{aligned}
	\end{equation}
	and therefore, 
	\begin{equation}
	\left[\left(E_{k}^{*} \otimes E_{j}\right) V\right]^{T}=\left[\begin{array}{ccccccc}
	0 & \cdots & 0 & \left(\rho_1^{\text{in}}\right)_{y v} & 0 & \cdots & 0 \\
	\vdots & & \vdots & \vdots & \vdots &  & \vdots \\
	0 & \cdots & 0 & \left(\rho_M^{\text{in}}\right)_{y v} & 0 & \cdots & 0
	\end{array}\right],
	\end{equation}
	where only the  $ (u-1)d+x $-th column is non-zero. Thus the  $ (j+(k-1)d^2) $-th column of $ B $ is
	\begin{equation}
	\left[0, \cdots, 0,  \left(\rho_1^{\text{in}}\right)_{y v}, \cdots, \left(\rho_M^{\text{in}}\right)_{y v}, 0, \cdots, 0\right]^{T}.
	\end{equation}
	The matrix $ I_{d^{2}} \otimes V^{T}=\operatorname{diag}\left(\left[V^{T}, \cdots, V^{T}\right]\right) $ is block diagonal.
	Thus  the $  (d^2\left[(u-1)d+x-1\right]+(v-1) d+y) $-th column of $ I_{d^{2}} \otimes V^{T} $ is the same as the $ (j+(k-1)d^2) $-th column of $ B $. Therefore, 
	\begin{equation}
	\left(I_{d^{2}} \otimes V^{T}\right) R=B,
	\end{equation}
	where $ R $ is a  $ d^4\times d^4 $ permutation matrix.
\end{IEEEproof}

Since we choose $ \{\sigma_{n}\}_{n=1}^{d^2} $ as the natural basis, we have
\begin{equation}\label{aa}
C=\left[\operatorname{vec}\left(P_{1}\right), \operatorname{vec}\left(P_{2}\right), \cdots, \operatorname{vec}\left(P_{L}\right)\right]^{T}.
\end{equation}
Using Proposition \ref{btensor} and \eqref{tensora}, we have
\begin{equation}\label{special}
\left(I_{M} \otimes C\right)K\left(I_{d^{2}} \otimes V^{T}\right) R\operatorname{vec}(X)=\operatorname{vec}({ \mathcal{P}^{T}}).
\end{equation}
Define $ Y\triangleq \left(I_{M} \otimes C\right)K\left(I_{d^{2}} \otimes V^{T}\right)R $. Since we assume the input states and measurement operators are both informationally complete or overcomplete,  the unique least squares solution to $\min || Y\text{vec}(\hat X)-\text{vec}(\widehat{ \mathcal{P}}^{T})||$ is 
\begin{equation}\label{ls}
\operatorname{vec}\left(\hat{X}_\text{LS}^{(1)}\right)=\left(Y^{\dagger} Y\right)^{-1} Y^{\dagger} \operatorname{vec}(\widehat{ \mathcal{P}}^{T}).
\end{equation}
However, the computational complexity for this least squares solution is usually $ O(d^{12}) $ \cite{cptp2018} which is quite high. To reduce the computational complexity, we do not actually utilize \eqref{ls}. Noting the special structure property of \eqref{special}, we firstly reconstruct $ \hat A $ and then obtain $ \hat X $, which is more efficient in computation.
To reconstruct $ \hat A $ from \eqref{maineq2}, we have 
\begin{equation}\label{lsl}
\operatorname{vec}(\hat{A})=K^{T}\left(I_{M} \otimes\left(C^{\dagger} C\right)^{-1} C^{\dagger}\right) \operatorname{vec}(\widehat{ \mathcal{P}}^{T}),
\end{equation}
and then the least squares solution $\operatorname{vec}\left(\hat{X}_\text{LS}^{(2)}\right)$ is given by
\begin{equation}\label{simls}
\begin{aligned}
\operatorname{vec}\left(\hat{X}_\text{LS}^{(2)}\right)&=\left(B^{\dagger} B\right)^{-1} B^{\dagger} \operatorname{vec}(\hat{A}) \\
&=R^{T}\left(I_{d^2} \otimes\left(V^{*} V^{T}\right)^{-1} V^{*}\right) \operatorname{vec}(\hat{A}).
\end{aligned}
\end{equation}
Therefore, we can directly obtain the least squares
 solution $\operatorname{vec}\left(\hat{X}_\text{LS}^{(2)}\right)$ as
\begin{equation}\label{tls}
\begin{aligned}
\operatorname{vec}\left(\hat{X}_\text{LS}^{(2)}\right)=&R^{T}\left(I_{d^{2}} \otimes\left(V^{*} V^{T}\right)^{-1} V^{*}\right) K^{T}\left(I_{M} \otimes\left(C^{\dagger} C\right)^{-1} C^{\dagger}\right) \operatorname{vec}(\widehat{ \mathcal{P}}^{T}).
\end{aligned}
\end{equation}

\begin{remark}
	The procedure to obtain $\hat A  $ in \eqref{lsl} is usually known as QST for output states \cite{qci,8022944} but our method
	is  different from such a procedure. Here we do not consider any constraints, i.e., Hermitian, positive semidefinite and unit trace, on the estimate $ \hat\rho_{m}^\text{out} $ and we directly reconstruct the process from the data vector by linear inversion.	
	 Of course, it is possible to obtain an alternative physical estimate from  $ \hat\rho_{m}^\text{out} $ as in \cite{effqst} or via Maximum Likelihood Estimation (MLE). However, here $\hat{A}$ is unnecessary to be  a physical estimate because it
	is only an intermediate product, and our direct method can already guarantee the ultimate process estimation to be physical by later solving Problem \ref{subproblem1} and Problem \ref{subproblem2} in Sec. \ref{twostage}.
\end{remark}

Considering the positive semidefinite requirement $ X\geq 0 $ and the  constraint  $\operatorname{Tr}_1(X)\leq I_d$, we convert  QPT 
into an optimization problem as follows.
\begin{problem}\label{problem2}
	Given the parameterization matrix $V$ of all the input states, the permutation matrix $ R $ and reconstructed $\hat A$, find a Hermitian and positive semidefinite estimate $\hat{X}$ minimizing
	$$\left\|\hat X-\operatorname{vec}^{-1}\left(R^{T}\left(I_{d^2} \otimes\left(V^{*} V^{T}\right)^{-1} V^{*}\right) \operatorname{vec}(\hat{A})\right)\right\|,$$ 
	such that $\operatorname{Tr}_1(\hat X)\leq I_d$.
\end{problem}

$  $
\subsection{Two-stage solution of QPT}\label{twostage}
The direct solution to Problem \ref{problem2} is usually difficult  and we split it into two   optimization sub-problems  (i.e., a two-stage solution):
\addtocounter{problem}{-1}
\renewcommand{\theproblem}{\arabic{problem}$.1$}
\begin{problem}\label{subproblem1}
	Let $$\hat D\triangleq\hat{X}_\text{LS}^{(2)}=\operatorname{vec}^{-1}\left(R^{T}\left(I_{d^2} \otimes\left(V^{*} V^{T}\right)^{-1} V^{*}\right) \operatorname{vec}(\hat{A})\right)$$ be a given matrix. Find a Hermitian and positive semidefinite $d^2\times d^2$ matrix $\hat G$ minimizing $||\hat G-\hat D||$.
\end{problem}

\renewcommand{\theproblem}{\arabic{problem}}
\addtocounter{problem}{-1}
\renewcommand{\theproblem}{\arabic{problem}$.2$}
\begin{problem}\label{subproblem2}
	Let $\hat G\geq 0$ be given.  Find a Hermitian and positive semidefinite $d^2\times d^2$ matrix $\hat X$ minimizing $||\hat X-\hat G||$,
	such that $\operatorname{Tr}_1(\hat X)\leq I_d$.
\end{problem}
\renewcommand{\theproblem}{\arabic{problem}}

For Problem \ref{subproblem1}, note that
if $ S=S^{\dagger}, W=-W^{\dagger} $, we have $\|S+W\|^{2}=\|S\|^{2}+\|W\|^{2}$. Therefore,
\begin{equation}
\|\hat G-\hat D\|^{2}=\left\|\hat G- \frac{\hat D+\hat D^{\dagger}}{2}\right\|^{2}+\left\| \frac{\hat D-\hat D^{\dagger}}{2}\right\|^{2}.
\end{equation}
We perform the spectral decomposition as $\frac{\hat D+\hat D^{\dagger}}{2}=U\hat{K}U^{\dagger}  $  where $\hat{K}=\operatorname{diag}\left(k_{1}, \cdots, k_{d^{2}}\right)$ is a  diagonal matrix. We define $ \hat Z\triangleq U^{\dagger} \hat GU $. Since $ \hat G $ is positive semidefinite, we have $  \hat{Z}_{ii}\geq0 $. Therefore,
\begin{align*}
\left\|\hat G- \frac{\hat D+\hat D^{\dagger}}{2}\right\|^{2} &=\|\hat K -\hat Z\|^{2} \\
&=\sum_{i \neq j} \left|\hat{Z}_{ij}\right|^{2}+\sum_{i}\left(k_{i}-\hat{Z}_{ii}\right)^{2} \\
& \geq \sum_{k_{i}<0}\left(k_{i}-\hat{Z}_{ii}\right)^{2}\\
& \geq \sum_{k_{i}<0} k_{i}^{2}. \numberthis
\end{align*}
Therefore, the unique optimal solution is $ \hat{Z}=\operatorname{diag}(b) $ where
\begin{equation}
b_{i}= \begin{cases}
k_{i}, & k_{i} \geq 0, \\ 
0, & k_{i}<0,
\end{cases}
\end{equation}
and $ \hat G=U\operatorname{diag}(b)U^{\dagger} $.

After solving Problem \ref{subproblem1}, we rewrite the spectral decomposition of $ \hat G$ as
\begin{equation}\label{geig}
\begin{aligned}
&\hat{G}=\sum_{i=1}^{d^2} \operatorname{vec}\left(\hat{S}_{i}\right) \operatorname{vec}\left(\hat{S}_{i}\right)^{\dagger},\\
\end{aligned}
\end{equation}
where each $ \hat{S}_{i} \in \mathbb{C}^{d\times d}$. Define $\hat F\triangleq\sum_{i=1}^{d^{2}} \hat{S}_{i} \hat{S}_{i}^{\dagger}$.  Using \eqref{property4}, we have
\begin{equation}\label{trg}
\begin{aligned}
\operatorname{Tr}_{1}(\hat{G})&=\operatorname{Tr}_{1}\left(\sum_{i=1}^{d^{2}} \operatorname{vec}\left(\hat{S}_{i}\right) \operatorname{vec}\left(\hat{S}_{i}\right)^{\dagger}\right)\\
&=\sum_{i=1}^{d^{2}} \hat{S}_{i} \hat{S}_{i}^{\dagger}=\hat F.
\end{aligned}
\end{equation}
Then, we consider the partial trace constraint by correcting $\hat{G}$.
Assume that  the spectral decomposition of $ \operatorname{Tr}_{1} (X)=F $  is
\begin{equation}\label{ff}
F=U_{F} \operatorname{diag}\left(f_{1}, \cdots, f_{d}\right) U_{F}^{\dagger},
\end{equation}
where $ 1\geq f_{1} \geq\cdots f_{d} \geq 0$ and the spectral decomposition of $ \hat F $ is
\begin{equation}\label{hatf}
\hat{F}=U_{\hat F} \operatorname{diag}\left(\hat f_{1}, \cdots, \hat f_{d}\right) U_{\hat F}^{\dagger},
\end{equation}
where $ \hat f_{1} \geq\cdots \geq\hat f_{c}>0$ and $ \hat f_{c+1}= \cdots \hat f_{d}=0 $, i.e., $ \operatorname{Rank}(\hat F)=c $. 
Then, we define
\begin{equation}\label{barf}
\bar{F}\triangleq U_{\hat F} \operatorname{diag}\left(\bar f_{1}, \cdots, \bar f_{d}\right) U_{\hat F}^{\dagger},
\end{equation}
where $ \bar f_{i}=\hat f_{i} $ for $ 1\leq i \leq c $, $ \bar f_{i}=\frac{\hat f_{c}}{N} $ for $ c+1\leq i \leq d$, and $ N $ is the number of copies for each input states. Since $ \bar{F} $ is invertible, $ \bar{F}^{-1/2} $ is well defined and we also define
\begin{equation}
\tilde{F}\triangleq U_{\hat F} \operatorname{diag}\left(\tilde{f}_{1}, \cdots, \tilde{f}_{d}\right) U_{\hat F}^{\dagger},
\end{equation}
where $ \tilde{f}_{i}=\min\left({\bar f}_{i},1\right) $ for $ 1\leq i\leq d $. Thus, $ \tilde{f}_{i} \leq 1 $ for $ 1\leq i \leq d $.
To solve Problem \ref{subproblem2} and satisfy the partial trace constraint, let $ \hat Q_{i}\triangleq\tilde F^{1/2}\bar F^{-1/2}\hat S_{i} $.
We then obtain  $ \hat X $ as
\begin{equation}\label{xeig}
\begin{aligned}
&\hat{X}=\sum_{i=1}^{d^{2}} \operatorname{vec}\left(\hat{Q}_{i}\right) \operatorname{vec}\left(\hat{Q}_{i}\right)^{\dagger},
\end{aligned}
\end{equation}
where $  \left\{\operatorname{vec}\left(\hat{Q}_{i}\right)\right\}_{i=1}^{d^2} $ might not be an orthogonal basis. From \eqref{xeig}, $ \hat X $ is Hermitian, positive semidefinite and also satisfies 
\begin{align*}
\operatorname{Tr}_{1}(\hat{X})&=\operatorname{Tr}_{1}\left(\sum_{i=1}^{d^{2}} \operatorname{vec}\left(\hat{Q}_{i}\right) \operatorname{vec}\left(\hat{Q}_{i}\right)^{\dagger}\right)\\
&=\sum_{i=1}^{d^{2}} \hat{Q}_{i} \hat{Q}_{i}^{\dagger}=\tilde F^{1/2}\bar{F}^{-1 / 2} \sum_{i=1}^{d^{2}} \hat{S}_{i} \hat{S}_{i}^{\dagger} \bar{F}^{-1 / 2}\tilde F^{1/2}\\
&=\tilde F^{1/2}\bar{F}^{-1 / 2} \hat F \bar{F}^{-1 / 2}\tilde F^{1/2}\\
&=\tilde F^{1/2}U_{\hat F}\operatorname{diag}\left(1,\cdots,1,0,\cdots, 0\right)U_{\hat F}^{\dagger}\tilde F^{1/2}\\
&=U_{\hat F}\operatorname{diag}\left(\tilde{f}_{1},\cdots,\tilde{f}_{c},0,\cdots, 0\right)U_{\hat F}^{\dagger} \\
&\leq I_{d}. \numberthis \label{hatx}
\end{align*}

The Stage 1 solution $ \hat G $ is a projection
under the Frobenius norm  satisfying the CP constraint, which was also discussed in \cite{cptp2018,surawy2021projected}. The TP constraint  makes QPT a nontrivial extension of
state tomography \cite{cptp2018}. The work in \cite{cptp2018,surawy2021projected} introduced two iterative algorithms designed to ensure that the estimate satisfies the CPTP constraint. While these algorithms may offer good accuracy, our two-stage algorithm can achieve a closed-form solution. We can also efficiently optimize the input states and the measurement operators using the TSS algorithm. It is worth highlighting that our algorithm is versatile and can be applied to non-TP cases.  Additionally, considering Hamiltonian identification in \cite{8022944} where  $ \operatorname{Rank}(\hat X)=1 $ and $\operatorname{Tr}_1(\hat X)= I_d$, our Stage 2 solution is  optimal for Problem \ref{subproblem2} which has been proved in  \cite{8022944}. For general QPT,  the estimated process matrix $ \hat X $ is a sub-optimal solution.

\begin{remark}
	If we have the prior knowledge that the quantum process is TP, we can assume that $ \hat F $ is non-singular (i.e., $\hat F >0  $) and it is true when the  number of copies is sufficiently large, because $ \hat F $  converges to $ I_d $ as $ N $ tends to infinity. Thus, we correct $ \hat Q_{i} $ as $ \hat Q_{i}=\hat F^{-1/2}\hat S_{i} $ and $ \hat{X}=\sum_{i=1}^{d^{2}} \operatorname{vec}\left(\hat{Q}_{i}\right) \operatorname{vec}\left(\hat{Q}_{i}\right)^{\dagger} $ where $ \operatorname{Tr}_{1}(\hat{X})=I_d $.
\end{remark}

\section{Computational complexity and storage requirements}\label{sec4}
\begin{figure}
	\centering
	\includegraphics[width=4.5in]{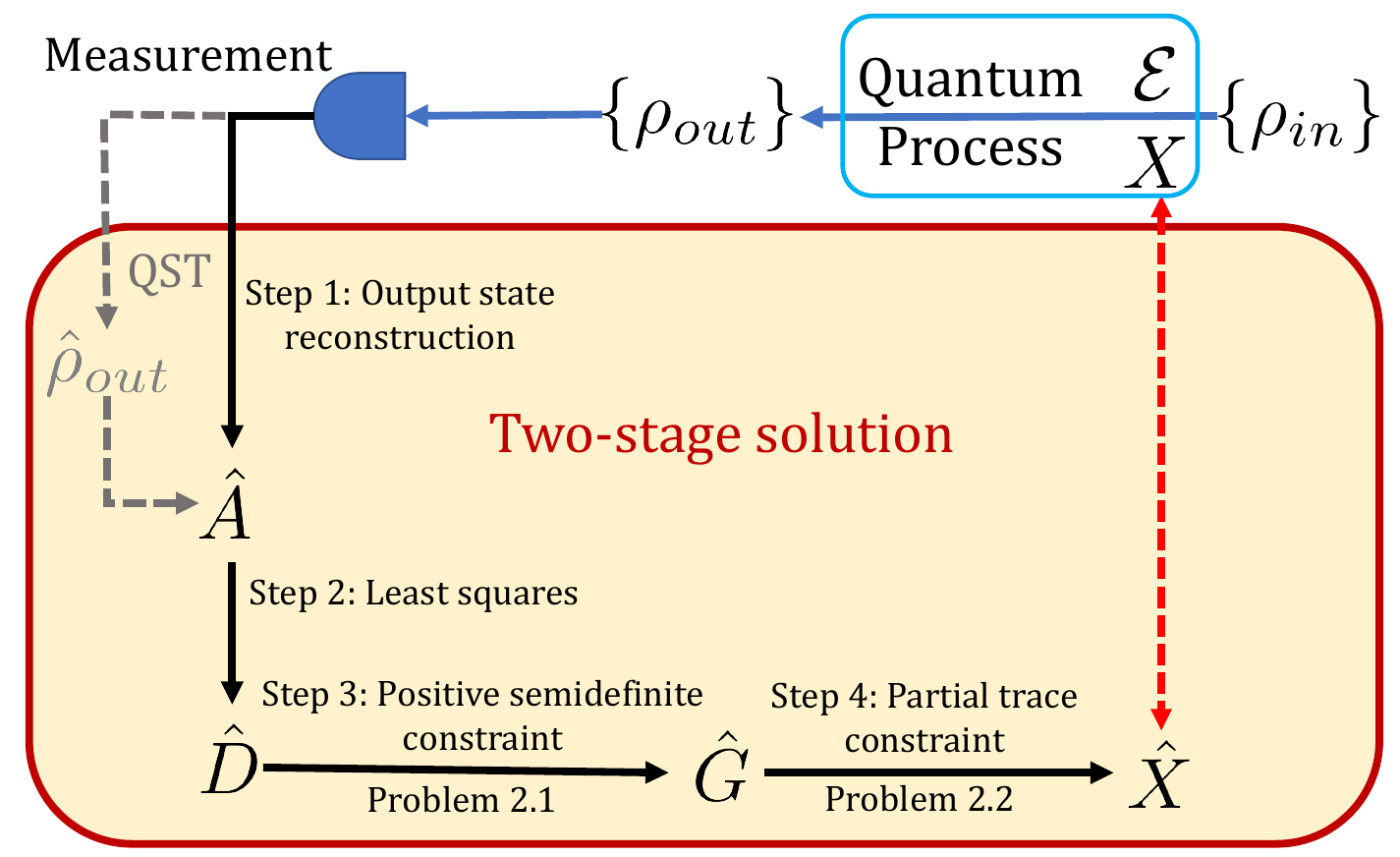}
	\centering{\caption{Procedure for our TSS algorithm for QPT which has four steps. In Step 1, using \eqref{lsl}, we reconstruct the parameterization matrix of all of the output states $ \hat A $ directly from measurement data  which is different from the QST in \cite{qci,8022944}. Here $\hat{A}$ is not needed be  a physical estimate because it is only an intermediate product. In Step 2, we utilize \eqref{simls}. Step 3 and Step 4 address the positive semidefinite constraint and partial trace constraint, respectively, by solving Problems \ref{subproblem1} and \ref{subproblem2}. These two steps together constitute our two-stage solution.}\label{procedure}}
\end{figure}
We summarize our  procedure for the QPT framework  with the TSS algorithm in Fig. \ref{procedure}.
In this paper,  we do not consider the time spent on experiments. Here we discuss the computational
complexity and storage requirement in each step in the red box of Fig. \ref{procedure} and give the total time and space complexity in Table \ref{table1}.

\textbf{{Step 1}}.  We reconstruct $\hat A $ using  \eqref{lsl}. The computational complexity is $ O(Ld^4) $ for $ \left(C^{\dagger} C\right)^{-1} C^{\dagger} $, $ O(MLd^2) $ for $ \left(I_{M} \otimes \left(C^{\dagger} C\right)^{-1} C^{\dagger}\right) \operatorname{vec}(\widehat{ \mathcal{P}}^{T}) $ and $ O(Md^2) $ for $ K^{T}\left(I_{M} \otimes \left(C^{\dagger} C\right)^{-1} C^{\dagger}\right) \operatorname{vec}(\widehat{ \mathcal{P}}^{T}) $ where $ M \geq d^2 $ is the number of different  input states and 
$ L\geq d^2$ is the number of different  measurement operators.
Thus, the total computational complexity is $ O(MLd^2) $.
For storage requirement, the storage is $ O(Ld^2) $ for the parameterization matrix $ C $ of measurement operators, $ O(ML) $ for measurement data $ \widehat{ \mathcal{P}}$ and $ O(Md^2) $ for $\hat A$. Since $ K $ is a permutation matrix, it can be stored with an expense $ O(Md^2) $.

\textbf{Step 2}. Using \eqref{simls}, the computational complexity is $ O(Md^4) $ for $ \left(V^{*} V^{T}\right)^{-1} V^{*} $, $ O(Md^4) $ for\\ $ \left(I_{d^2} \otimes \left(V^{*} V^{T}\right)^{-1} V^{*}\right) \operatorname{vec}(\hat{A}) $  and $ O(d^4) $ for $ R^{T}\left(I_{d^2} \otimes\left(V^{*} V^{T}\right)^{-1} V^{*}\right) \operatorname{vec}(\hat{A}) $. Thus, the total computational complexity is $ O(Md^4) $. We need to store $ \left(V^{*} V^{T}\right)^{-1} V^{*} $  and $ R^{T} $. Since $ \left(V^{*} V^{T}\right)^{-1} V^{*} $ is a $ d^2\times M$ matrix and $ R $ is a permutation matrix, the storage requirements are $ O(Md^2) $ and $ O(d^2) $, respectively.

\textbf{Step 3}. To solve Problem \ref{subproblem1}, the computational complexity is determined by the spectral decomposition of $  \frac{\hat D+\hat D^{\dagger}}{2}$, which is $O(d^6)$. The storage requirements  are $ O(d^4) $ for $\hat D$, $ O(d^4)  $ for the spectral decomposition of $  \frac{\hat D+\hat D^{\dagger}}{2}$, and $ O(d^4)  $ for $\hat G$.

\textbf{Step 4}.  To solve Problem \ref{subproblem2}, the computational complexity for calculating $ \hat F $, $ \bar F^{-1/2} $ and  $ \tilde F^{1/2} $ are $ O(d^4) $, $O(d^3)$ and $O(d^3)$, respectively. Then the computational complexity for calculating $ \hat X $  is $O(d^6)$.  The storage requirements for $ \{\hat S_i\}_{i=1}^{d^2} $ and $ \{\hat Q_i\}_{i=1}^{d^2} $ are both  $ O(d^4) $. Then the storage for $\hat F$, $ \bar F $ and $ \tilde F $ are all $ O(d^2) $, and for $ \hat X $ is $ O(d^4) $.

\begin{table}[H]
	\caption{Computational complexity and storage requirement}
	\renewcommand{\arraystretch}{1.4}
	\label{table1} 
	
	\centering

	\begin{tabular}{c|c|c}
		\toprule
		&Computational complexity&Storage requirement\\
		\hline
		Step 1& $ O(MLd^2) $ & $ O(ML) $\\
		\hline
		Step 2& $ O(Md^4) $ &$ O(Md^2) $\\
		\hline
		Step 3& $O(d^6)$ &$ O(d^4) $\\
		\hline
		Step 4& $O(d^6)$ &$ O(d^4) $\\
		\hline
		Total& $ O(MLd^2) $  & $ O(ML) $\\
		\bottomrule
	\end{tabular}
	
\end{table}

The total computational complexity and  storage requirements for our TSS algorithm are presented in Table \ref{table1}.
Since $ M\geq d^2 $ and $ L\geq d^2 $, the total computational complexity is $ O(MLd^2) $. From \cite{qci,qptre,cptp2018}, the computational complexity for a direct least squares solution with $ d^2 $ different input states is $ O(d^{12}) $, which is significantly higher than $O(MLd^2)  $. The reason we obtain a lower computational complexity here is that 
we utilize the special structure of $B$ described in Proposition \ref{btensor}.
As for the total storage requirement, our TSS algorithm  is $ O(ML) $.
Without employing the tensor structure as  \eqref{simls}, the storage space will be $ O(Md^6) $ because  the storage requirement of $ B $ will be $ O(Md^6) $.

\section{Error analysis}\label{sec5}
From Fig. \ref{procedure}, our TSS algorithm has four steps. In this section, we analyze the error upper bound and 
 give the following theorem to characterize the  error upper bound analytically.
\begin{theorem}\label{mainthe}
	If $\{E_i\}_{i=1}^{d^2}$ and $\{\sigma_n\}_{n=1}^{d^2}$ are chosen as the natural basis and the input quantum states are $\{\rho_m^{\text{in}}\}_{m=1}^{M}$, then the estimation error of the TSS algorithm $\mathbb E||\hat X-X||$ scales as $$ O\!\!\left(\!\!\frac{\sqrt{d}\operatorname{Tr}({F}) \sqrt{J\operatorname{Tr}\left(\left(C^{\dagger} C\right)^{-1}\right)}\!\sqrt{M\operatorname{Tr}\left(\left(V^{*} V^{T}\right)^{-1}\right)}}{\sqrt{N}}\!\right), $$ where $N$ is the number of copies for each output state, $ J $ is the number of POVM sets, $ C $ is defined as \eqref{aa}, $ V $ is defined as \eqref{probestate}, $ F=\operatorname{Tr}_{1} (X) $ and $\mathbb E(\cdot)$ denotes expectation with respect to all possible measurement results.
\end{theorem}
\begin{IEEEproof}
	 We will first calculate the error for each of the four steps  as shown in Fig. \ref{procedure} and then present the final error bound. The main tools in this proof include the error analysis of QST in \cite{Qi2013} and some matrix inequalities in Appendix \ref{appendixa}. For any quantity $ S $, we denote its estimation error as $ \Delta_{S}\triangleq ||\hat S- S|| $.
	\subsection{{Error in Step 1}} 
	The MSE of the $ m $-th estimated output state, i.e.,  $  \operatorname{col}_{m}(\hat A^{T})$ is asymptotically (see \cite{Qi2013}) 
	\begin{equation}\label{qstmse}
	\begin{aligned}
	&\mathbb E\left\|\operatorname{col}_{m}(\hat A^{T})-\operatorname{col}_{m}( A^{T})\right\|^{2}\\
	=&\frac{J}{N} \operatorname{Tr}\left(\left(C^{\dagger} C\right)^{-1} C^{\dagger} \mathrm P C\left(C^{\dagger} C\right)^{-1}\right),
	\end{aligned}
	\end{equation}
	where $N$ is the number of copies for each input state and $$\mathrm P\triangleq\operatorname{diag}\left(p_{m1}-p_{m1}^{2}, \cdots, p_{mL}-p_{mL}^{2}\right).$$ Therefore, the MSE is bounded by \cite{Qi2013}
	\begin{equation}\label{stateupper}
	\mathbb E\left\|\operatorname{col}_{m}(\hat A^{T})-\operatorname{col}_{m}( A^{T})\right\|^{2}\leq\frac{J}{4N}\operatorname{Tr}\left(\!\left(C^{\dagger} C\right)^{-1} \right).
	\end{equation}
	Then, for $ \hat A $, we have
	\begin{equation}\label{lambupper}
	\begin{array}{rl}
	\mathbb E\Delta_A^{2} &=\sum_{m=1}^{M} \mathbb E\left\|\operatorname{col}_{m}(\hat A^{T})-\operatorname{col}_{m}( A^{T})\right\|^{2}\\
	&\leq \dfrac{MJ}{4N}\operatorname{Tr}\left(\left(C^{\dagger} C\right)^{-1} \right).
	\end{array}
	\end{equation}

	\subsection{Error in Step 2}

	Since 
	\begin{equation}\label{vv}
	\begin{aligned}
	\left\|\left(V^{*} V^{T}\right)^{-1} V^{*}\right\|^{2}&=\operatorname{Tr}\left(V^{T}\left(V^{*} V^{T}\right)^{-1}\left(V^{*} V^{T}\right)^{-1} V^{*}\right)\\
	&=\operatorname{Tr}\left(\left(V^{*} V^{T}\right)^{-1}\right),
	\end{aligned}
	\end{equation}
	we have
	\begin{align*}
	&\quad\Delta_{D}^2=\|\hat{D}-D\|^{2}\\
	&=\left\|\operatorname{vec}\left(\hat D\right)-\operatorname{vec}\left( D\right)\right\|^2 \\
	&=\left\|R^{T}\left(I \otimes\left(V^{*} V^{T}\right)^{-1} V^{*}\right)(\operatorname{vec}(\hat{A})-\operatorname{vec}(A))\right\|^{2} \\
	&\leq\left\|\left(V^{*} V^{T}\right)^{-1} V^{*}\right\|^{2}\|\operatorname{vec}(\hat{A})-\operatorname{vec}(A)\|^{2}\\
	&=\operatorname{Tr}\left(\left(V^{*} V^{T}\right)^{-1}\right)\Delta_A^2, \numberthis \label{dest}
	\end{align*}
	where we use Proposition \ref{proposition2} for the inequality.
	\subsection{Error in Step 3}
	Since  $\hat G$ minimizes $||\hat G-\hat D||$, we have $ \|\hat{G}-\hat{D}\|\leq\|\hat{D}-D\| $ and thus
	\begin{equation}\label{ggerror}
	\begin{aligned}
	\Delta_{G}=&\|\hat{G}-G\|=\|\hat{G}-D\| \leq\|\hat{G}-\hat{D}\|+\|\hat{D}-D\| \leq 2\Delta_{D},
	\end{aligned}
	\end{equation}
	because the true value $ G=D $.
	\subsection{Error in Step 4}
	For $ \hat F $, if $ 1\geq \hat f_1 \geq \cdots \geq \hat f_d \geq0$, i.e., $ \hat F \leq I $, we have $ \tilde{F}^{1 / 2} \bar{F}^{-1 / 2}=I $ and thus $ \hat X=\hat G $. 
	Therefore, using \eqref{lambupper}, \eqref{dest} and \eqref{ggerror}, we obtain the final error bound as
	\begin{equation}\label{finalcase1}
	\begin{aligned}
 \mathbb E\|\hat{X}-X\|= \mathbb E\|\hat{G}-G\| \sim O\left(\frac{ \sqrt{J\operatorname{Tr}\left(\left(C^{\dagger} C\right)^{-1}\right)}\sqrt{M\operatorname{Tr}\left(\left(V^{*} V^{T}\right)^{-1}\right)}}{\sqrt{N}}\right).
	\end{aligned}
	\end{equation}
	
	Then we consider the case where at least one eigenvalue of $ \hat F $ is larger than $ 1 $. 
	Assume $ \hat{f}_{i}>1 $ for $ 1\leq i\leq a$ and $ 0\leq\hat{f}_{i}\leq1 $ for  $ a+1\leq i\leq d$ in \eqref{hatf}.
	From Lemma \ref{lemma2} and \eqref{trg}, we have 
	\begin{equation}\label{fi}
	\begin{aligned}
	\|\hat{F}-F\|&=\left\|\operatorname{Tr}_{1} (\hat{G})-\operatorname{Tr}_{1} (G)\right\|\\
	& \leq \sqrt{d}\|\hat{G}-G\|\\
	&=\sqrt{d}\Delta_{G}.
	\end{aligned}
	\end{equation}
	Using \eqref{ff}, \eqref{hatf}, and \eqref{weyl} in Appendix \ref{appendixa}, we can obtain
	\begin{equation}
	\max _{i}\left|\hat{f}_{i}-f_{i}\right| \leq \|\hat F-F\| \leq \sqrt{d} \Delta_{G},
	\end{equation}
	and therefore we have
	\begin{equation}\label{tracehatf}
	\operatorname{Tr}({F})-d \sqrt{d} \Delta_{G} \leq \operatorname{Tr}(\hat{F}) \leq \operatorname{Tr}({F})+d \sqrt{d} \Delta_{G}.
	\end{equation}
	Using \eqref{barf} and \eqref{weyl2} in Appendix \ref{appendixa}, we have
	\begin{align*}
	\|\tilde{F}-\hat{F}\|^{2}&=\sum_{i=1}^{d}\left(\tilde{f}_{i}-\hat{f}_{i}\right)^{2}\\
	&=
	\sum_{i=1}^{a}\left(1-\hat{f_{i}}\right)^{2}+\sum_{i=c+1}^{d}\left(\frac{\hat f_c}{N}\right)^{2}\\
	& \leq \sum_{i=1}^{d}\left(f_{i}-\hat{f_{i}}\right)^{2}+(d-c)\left(\frac{\hat f_c}{N}\right)^{2}\\
	&\leq \|\hat{F}-{F}\|^{2}+O\left(\frac{1}{N^2}\right). \numberthis \label{tracetildef}
	\end{align*}
	Since $\lim _{N \rightarrow \infty} \Delta_{G}=0$, we have  $\operatorname{Tr} (\hat{G})=\operatorname{Tr}(\hat{F})\sim \operatorname{Tr}({F})$, and thus \eqref{tracetildef} indicates
	\begin{equation}\label{trace}
	\operatorname{Tr}(\hat{F})\sim \operatorname{Tr}(\tilde{F}) \sim \operatorname{Tr}({F}).
	\end{equation}
	
	Then we consider $ \tilde F^{1 / 2} \bar F^{-1 / 2} $ as
	\begin{equation}
	\tilde{F}^{1 / 2} \bar{F}^{-1 / 2}=U_{\hat{F}} \operatorname{diag}\left(\sqrt{\frac{\tilde{f}_{1}}{\bar{f}_{1}}}, \cdots \sqrt{\frac{\tilde{f}_{d}}{\bar{f}_{d}}}\right) U_{\hat{F}}^{\dagger}.
	\end{equation}
	For $ a+1 \leq i \leq d $, $ \tilde f_i =\bar f_i $. For $ 1\leq i\leq a$, $ \bar f_i=\hat f_i $, and we have
	\begin{equation}\label{fsq}
	\sqrt{\frac{\tilde{f}_{i}}{\bar{f}_{i}}}-1=\sqrt{\frac{1}{1+\left(\hat{f_{i}}-1\right)}}-1=\frac{1-\hat{f}_{i}}{2}+o\left(\frac{\hat{f_{i}}-1}{2}\right).
	\end{equation}
	Thus, using \eqref{fsq} and \eqref{weyl2} in Appendix \ref{appendixa}, we have
	\begin{equation}
	\begin{aligned}
	&\quad\left\|\tilde{F}^{1 / 2} \bar{F}^{-1 / 2}-I_{d}\right\|^{2}=\sum_{i=1}^{a}\left(\frac{1-\hat{f}_{i}}{2}+o\left(\frac{1-\hat{f}_{i}}{2}\right)\right)^{2} \\
	&\sim \frac{\sum_{i=1}^{a}\left(\hat{f}_{i}-1\right)^{2}}{4} \leq
	\frac{\sum_{i=1}^{d}\left(\hat{f_{i}}-f_{i}\right)^{2}}{4}\leq \frac{\|\hat{F}-{F}\|^{2}}{4}.
	\end{aligned}
	\end{equation}
	
	Then using \eqref{hatx} and \eqref{trace}, we have
	\begin{equation}
	\begin{aligned}
	\operatorname{Tr}(\hat{X})&=\operatorname{Tr}\left(\operatorname{Tr}_{1} (\hat X)\right)=\sum_{i=1}^{c} \tilde{f}_i\\
	&=\operatorname{Tr}(\tilde{F})-\frac{(d-c)\hat{f}_{c}}{N}\sim\operatorname{Tr}({F}).
	\end{aligned}
	\end{equation}
	Using \eqref{xeig}, we thus have
	\begin{equation}\label{normq}
	\begin{aligned}
	\sum_{i=1}^{d^{2}}\left\|\operatorname{vec}\left(\hat{Q}_{i}\right)\right\|^{2}&=\sum_{i=1}^{d^{2}} \operatorname{vec}\left(\hat{Q}_{i}\right)^{\dagger} \operatorname{vec}\left(\hat{Q}_{i}\right) \\
	&=\operatorname{Tr}(\hat{X})\sim \operatorname{Tr}({F}).
	\end{aligned}
	\end{equation}
	Similarly, using \eqref{geig} and \eqref{trg}, we have
	\begin{equation}\label{norms}
	\begin{aligned}
	\sum_{i=1}^{d^{2}}\left\|\operatorname{vec}\left(\hat{S}_{i}\right)\right\|^{2}&=\sum_{i=1}^{d^{2}} \operatorname{vec}\left(\hat{S}_{i}\right)^{\dagger} \operatorname{vec}\left(\hat{S}_{i}\right)\\
	&=\operatorname{Tr}(\hat{G})=\operatorname{Tr}\left(\operatorname{Tr}_{1} (\hat G)\right)\\
	&=\operatorname{Tr}(\hat F) \sim \operatorname{Tr}( F).
	\end{aligned}
	\end{equation}
	Thus, from  \eqref{property2} and the definition of $ \hat Q_i $, we know $\operatorname{vec}\left(\hat{Q}_{i}\right)= \left( I_{d} \otimes  \tilde F^{1 / 2} \bar F^{-1 / 2}\right)\operatorname{vec}\left(\hat{S}_{i}\right) $, and we have
	\begin{align*}
	&\quad\sum_{i=1}^{d^{2}}\left\|\operatorname{vec}\left(\hat{Q}_{i}\right)-\operatorname{vec}\left(\hat{S}_{i}\right)\right\|^{2}\\
	&=\sum_{i=1}^{d^{2}}\left\|\left( I_{d} \otimes \tilde F^{1 / 2} \bar F^{-1 / 2}-I_{d} \otimes I_{d}\right) \operatorname{vec}\left(\hat{S}_{i}\right)\right\|^{2} \\
	&\leq\left\|\tilde F^{1 / 2} \bar F^{-1 / 2}-I_{d}\right\|^{2} \sum_{i=1}^{d^{2}}\left\|\operatorname{vec}\left(\hat{S}_{i}\right)\right\|^{2} \\
	&\sim\frac{\operatorname{Tr}({F})}{4}\|\hat F-F\|^{2} \sim \frac{d\operatorname{Tr}({F})}{4}\|G-\hat{G}\|^{2}, \numberthis \label{qs}
	\end{align*}
	where we use  Proposition \ref{proposition2} in the inequality. Therefore, using 
	\eqref{normq}--\eqref{qs}, we have
	\begin{align*}
	&\quad\|\hat{X}-\hat{G}\|^{2}\\
	&=\left\|\sum_{i=1}^{d^{2}}\left(\operatorname{vec}\left(\hat{Q}_{i}\right) \operatorname{vec}\left(\hat{Q}_{i}\right)^{\dagger}-\operatorname{vec}\left(\hat{S}_{i}\right) \operatorname{vec}\left(\hat{S}_{i}\right)^{\dagger}\right)\right\|^{2} \\
	&\leq\left(\sum_{i=1}^{d^{2}}\left\|\operatorname{vec}\left(\hat{Q}_{i}\right) \operatorname{vec}\left(\hat{Q}_{i}\right)^{\dagger}-\operatorname{vec}\left(\hat{S}_{i}\right) \operatorname{vec}\left(\hat{S}_{i}\right)^{\dagger}\right\|\right)^{2} \\
	&=\Bigg(\sum_{i=1}^{d^{2}}\Big\|\operatorname{vec}\left(\hat{Q}_{i}\right) \operatorname{vec}\left(\hat{Q}_{i}\right)^{\dagger}-\operatorname{vec}\left(\hat{S}_{i}\right) \operatorname{vec}\left(\hat{Q}_{i}\right)^{\dagger}+\operatorname{vec}\left(\hat{S}_{i}\right) \operatorname{vec}\left(\hat{Q}_{i}\right)^{\dagger}-\operatorname{vec}\left(\hat{S}_{i}\right) \operatorname{vec}\left(\hat{S}_{i}\right)^{\dagger}\Big\|\Bigg)^{2} \\
	&\leq\Bigg(\sum_{i=1}^{d^{2}}\left(\left\|\operatorname{vec}\left(\hat{Q}_{i}\right)^{\dagger}\right\|+\left\|\operatorname{vec}\left(\hat{S}_{i}\right)\right\|\right)\left\|\operatorname{vec}\left(\hat{Q}_{i}\right)-\operatorname{vec}\left(\hat{S}_{i}\right)\right\|\Bigg)^{2} \\
	&\leq\left(\sum_{i=1}^{d^{2}}\left(\left\|\operatorname{vec}\left(\hat{Q}_{i}\right)\right\|+\left\|\operatorname{vec}\left(\hat{S}_{i}\right)\right\|\right)^{2}\right)\cdot\left(\sum_{i=1}^{d^{2}}\left\|\operatorname{vec}\left(\hat{Q}_{i}\right)-\operatorname{vec}\left(\hat{S}_{i}\right)\right\|^{2}\right) \\
	&\leq 2\left(\sum_{i=1}^{d^{2}}\left\|\operatorname{vec}\left(\hat{Q}_{i}\right)\right\|^{2}+\sum_{i=1}^{d^{2}}\left\|\operatorname{vec}\left(\hat{S}_{i}\right)\right\|^{2}\right)\cdot\left(\sum_{i=1}^{d^{2}}\left\|\operatorname{vec}\left(\hat{Q}_{i}\right)-\operatorname{vec}\left(\hat{S}_{i}\right)\right\|^{2}\right)\\
	&\sim d\left(\operatorname{Tr}({F})\right)^2\|G-\hat{G}\|^{2}, \numberthis \label{xgerror}
	\end{align*}
	where we use the Cauchy-Schwarz inequality in the third inequality  and AM-QM inequality \cite{sedrakyan2018algebraic} in the fourth inequality.
	Thus, using \eqref{lambupper}, \eqref{dest}, \eqref{ggerror} and \eqref{xgerror}, the final error bound is
	\begin{equation}\label{final}
	\begin{aligned}
	&\quad \mathbb E\|\hat{X}-X\| \leq\mathbb E\|\hat{X}-\hat{G}\|+\mathbb E\|\hat{G}-G\|\\
	& \leq (\sqrt{d}\operatorname{Tr}({F})+1)\mathbb E\|\hat{G}-G\| \\
	&\leq O\left(\!\!\frac{\sqrt{d}\operatorname{Tr}({F}) \sqrt{J\operatorname{Tr}\left(\left(C^{\dagger} C\right)^{-1}\right)}\sqrt{M\operatorname{Tr}\left(\left(V^{*} V^{T}\right)^{-1}\right)}}{\sqrt{N}}\right).
	\end{aligned}
	\end{equation}

	Comparing \eqref{finalcase1} and \eqref{final}, we obtain the final error bound for any quantum process in the approximation sense as 
	\begin{equation}\label{finalall}
	\begin{aligned}
	\mathbb E\|\hat{X}-X\|	\leq O\left(\frac{\sqrt{d}\operatorname{Tr}({F}) \sqrt{J\operatorname{Tr}\left(\left(C^{\dagger} C\right)^{-1}\right)}\sqrt{M\operatorname{Tr}\left(\left(V^{*} V^{T}\right)^{-1}\right)}}{\sqrt{N}}\right),
	\end{aligned}
	\end{equation}	
	where $F=\operatorname{Tr}_{1} (X)$ represents the success probability of the quantum process \cite{PhysRevA.82.042307}.
\end{IEEEproof}

Note that when the quantum process is TP, $ \operatorname{Tr}(F)=d $ and the error analysis is similar to non-TP case. Then the final error bound is
\begin{equation}\label{tpbound}
\begin{aligned}
 \mathbb E\|\hat{X}-X\|\leq O\left(\frac{d^{3/2} \sqrt{J\operatorname{Tr}\left(\left(C^{\dagger} C\right)^{-1}\right)}\sqrt{M\operatorname{Tr}\left(\left(V^{*} V^{T}\right)^{-1}\right)}}{\sqrt{N}}\right).
\end{aligned}
\end{equation}
\section{{Optimization of input states and measurement operators}}\label{optimal}
To minimize the tomography error, the optimal  input  states $ \{\rho_m^{\text{in}}\}_{m=1}^{M} $ and optimal measurement operators  $ \{P_l\}_{l=1}^{L} $ are usually dependent on the specific process. One advantage of our TSS algorithm is that we can give an analytical error upper bound as \eqref{finalall} or \eqref{tpbound} which  depends on the input states and measurement operators instead of the unknown process. Therefore, we can optimize the input states and measurement operators based on the upper bound. In addition, we also consider how robust the estimated process $ \hat X $ is w.r.t. the measurement error, which is determined by \eqref{lsl} and \eqref{simls}.
Therefore, we use the condition numbers of $ I_{M} \otimes C $ and $ B $ to describe the robustness, which reflects the sensitivity of the solution to the perturbations in data. Moreover, we consider the relationship between $ M $ and the final estimation error with randomly generated input states.

\subsection{Optimal input quantum states}
Define the set with $ M $ different types of input states as
\begin{equation}
\begin{aligned}
\mathcal D(d,M) \triangleq&\Big\{\{\rho_m^{\text{in}}\}_{m=1}^{M} \mid \forall 1 \leq m \leq M, \rho_m^{\text{in}} \in \mathbb C^{d \times d}, \rho_m^{\text{in}}=\left(\rho_m^{\text{in}}\right)^{\dagger}, \rho_m^{\text{in}}\geq 0, \operatorname{Tr}(\rho_m^{\text{in}})=1\Big\}.
\end{aligned}
\end{equation}
Similarly to the optimal probe states in quantum detector tomography \cite{xiao2021optimal,xiaocdc2021}, we consider the optimal input states for QPT. The part of \eqref{finalall} or \eqref{tpbound} dependent on the input states is $ M\operatorname{Tr}\big((V^{*} V^{T})^{-1}\big) $, which will be our first cost function. Therefore, we define the set of all the optimal input states to minimize this bound  as 
\begin{equation}
\mathrm{OIS}_{1}(d,M) \triangleq \underset{\left\{\rho_m^{\text{in}}\right\}_{m=1}^{M} \in \mathcal D(d,M)}{\arg \min } M \operatorname{Tr}\left(\left(V^{*} V^{T}\right)^{-1}\right).
\end{equation}
For general QPT, we have no prior knowledge of the unknown process. From  \eqref{simls} with the reconstructed $ \hat A $, the following conditions are equivalent: (i) the process cannot be uniquely identified; (ii)  the input states do not span the space of all $d$-dimensional Hermitian matrices; (iii) $V^{*} V^{T}$ is singular; (iv)   $ M\operatorname{Tr}\big((V^{*} V^{T})^{-1}\big) $ is infinite. From (i) and (iv), it is thus reasonable to take  $ M\operatorname{Tr}\big((V^{*} V^{T})^{-1}\big) $ as a cost function.

In addition, maximizing the robustness of the reconstructed process with respect to measurement noise amounts to minimizing the condition number.
Therefore, from  \eqref{simls}, the second  cost function is the condition number of $ B $; i.e., $ \operatorname{cond}(B) $,  defined as $$ \operatorname{cond}(B)\triangleq\frac{\sigma_{\max }(B)}{\sigma_{\min }(B)} $$ and $\sigma_{\max/\min}(B)$ is the maximum/minimum singular value of $B$. Using \eqref{structureb}, we have $\operatorname{cond}(B)=\operatorname{cond}(I)\operatorname{cond}(V^{T})=\operatorname{cond}(V)$ which also only depends on the input states. Similarly, we also define the set of all the optimal input states to minimize the condition number as
\begin{equation}
\mathrm{OIS}_{2}(d,M) \triangleq \underset{\left\{\rho_m^{\text{in}}\right\}_{m=1}^{M} \in \mathcal D(d,M)}{\arg \min } \operatorname{cond}(V).
\end{equation}
 We then define two sets $ \underline{\mathrm{OIS}_{1}}(d,M) $ and $ \underline{\mathrm{OIS}_{2}}(d,M) $ to characterize two lower bounds of these two cost functions as
\begin{equation}
\begin{aligned}
\underline{\mathrm{OIS}_{1}}(d,M)\triangleq &\Big\{{\left\{\rho_m^{\text{in}}\right\}_{m=1}^{M}}\in \mathcal D(d,M) \mid M \operatorname{Tr}\left(\left(V^{*} V^{T}\right)^{-1}\right)=d^{4}+d^{3}-d^{2}\Big\},
\end{aligned}
\end{equation}
\begin{equation}
\underline{\mathrm{OIS}_{2}}(d,M) \triangleq \left\{{\left\{\rho_m^{\text{in}}\right\}_{m=1}^{M}}\in \mathcal D(d,M) \mid \operatorname{cond}(V)=\sqrt{d+1}\right\},
\end{equation}
and the proof is as given in Theorem \ref{theorem2}. 
In addition, we define the set of the optimal input states ($ \mathrm{OIS}(d,M) $) and the set of the optimal input states achieving both of the lower bounds ($\underline{\mathrm{OIS}}(d,M)$) as 
\begin{equation}
\begin{aligned}
\mathrm{OIS}(d,M)\triangleq \mathrm{OIS}_{1}(d,M)\cap \mathrm{OIS}_{2}(d,M),
\end{aligned}
\end{equation}
and
\begin{equation}
\begin{aligned}
\underline{\mathrm{OIS}}(d,M)\triangleq \underline{\mathrm{OIS}_{1}}(d,M) \cap \underline{\mathrm{OIS}_{2}}(d,M).
\end{aligned}
\end{equation}
Thus when $ \mathrm{OIS}(d,M) $ is non-empty,  it should minimize  $M\operatorname{Tr}\big((V^{*} V^{T})^{-1}\big)$ and minimize $\operatorname{cond}(V)$ \emph{simultaneously}. Additionally, when $ \underline{\mathrm{OIS}}(d,M) $ is non-empty, it can achieve the lower bounds of  $M\operatorname{Tr}\big((V^{*} V^{T})^{-1}\big)$ and $\operatorname{cond}(V)$ \emph{simultaneously}.
Then, we present the following theorem for $ \underline{\mathrm{OIS}}(d,M) $.
\begin{theorem}
	\label{theorem2}
	For a $ d $-dimensional quantum process, with $ M $ different types of input states and the parameterization matrix   $ V $ as in  \eqref{probestate}, we have $M\operatorname{Tr}\big((V^{*} V^{T})^{-1}\big)\geq d^{4}+d^{3}-d^{2}$ and $\operatorname{cond}(V)\geq \sqrt{d+1}$.
	These lower bounds are achieved simultaneously; i.e., $ {\left\{\rho_m^{\text{in}}\right\}_{m=1}^{M}}\in \underline{\mathrm{OIS}}(d,M) $ if and only if the eigenvalues of $ V^{*} V^{T}$     are  $ \tau_{1}=\frac{M}{d} $ and $\tau_{2}=\dots=\tau_{d^2}=\frac{M}{d(d+1)}  $.
\end{theorem}
\begin{IEEEproof}
	Denote  the eigenvalues of $ V^{*}V^{T} $ as $  \tau_{1} \geq \tau_{2} \geq \cdots \geq \tau_{d^2}>0  $. 
	Since  $ \left\{\rho_m^{\text{in}}\right\}_{m=1}^{M} \in \mathcal D(d,M) $,  	we can obtain two constraints $ \sum_{i=1}^{d^2} \tau_{i}\leq M$ and $\tau_{1}\geq \frac{M}{d} $ coming from the purity requirement and unit trace of quantum states, respectively, which have been proved in \cite{xiao2021optimal}. To minimize $M\operatorname{Tr}\big((V^{*} V^{T})^{-1}\big)$ and $\operatorname{cond}(V)$, we relax the optimization problems 
	as follows:
	\begin{equation}
	\label{umse}
	\begin{aligned}
	\text { min } & \sum_{i=1}^{d^2} \frac{M}{\tau_{i}} \\
	\text { s.t. } & \sum_{i=1}^{d^2} \tau_{i}\leq M,\tau_{1}\geq \frac{M}{d},
	\end{aligned}
	\end{equation}
	and
	\begin{equation}
	\label{robust}
	\begin{aligned}
	\text { min } & \sqrt{ \frac{\tau_{1}}{\tau_{d^2}}} \\
	\text { s.t. } & \sum_{i=1}^{d^2} \tau_{i}\leq M,\tau_{1}\geq \frac{M}{d}.
	\end{aligned}
	\end{equation}
	Using Lagrange multiplier method, we have
	\begin{equation}
	M\operatorname{Tr}\left(\left(V^{*} V^{T}\right)^{-1}\right)\geq d^{4}+d^{3}-d^{2},
	\end{equation}
	and
	\begin{equation}
	\operatorname{cond}(V) \geq	\sqrt{\frac{M / d}{M / \left(d(d+1)\right)}}=\sqrt{d+1}.
	\end{equation}
	These lower bounds hold if and only if  $ \tau_{1}=\frac{M}{d} $ and $\tau_{2}=\dots=\tau_{d^2}=\frac{M}{d(d+1)}  $.
\end{IEEEproof}

Since $  \mathcal D(d,M)  $ is a closed space, $ \text{OIS}_{1}(d,M) $ and $ \text{OIS}_{2}(d,M) $ are always non-empty for any $ M\geq d^2 $. However, it is not clear whether this also holds for $ \text{OIS}(d,M) $. For the lower bounds of these two cost functions, if one of them is achieved, we have $\tau_{1}= \frac{M}{d}, \tau_{2}=\dots=\tau_{d^2}=\frac{M}{d(d+1)} $ and thus 
\begin{equation}\label{oil}
\begin{aligned}
\underline{\mathrm{OIS}}(d,M)= \underline{\mathrm{OIS}_{1}}(d,M)= \underline{\mathrm{OIS}_{2}}(d,M).
\end{aligned}
\end{equation}
If the two lower bounds cannot be achieved, the above three sets are empty.
Therefore, \eqref{oil} always holds for any $ M\geq d^2 $. However, we still have not fully determined what values of $ M $ make \eqref{oil} non-empty. In addition, when $ \underline{\mathrm{OIS}}(d,M) $ is not empty, we have $\underline{\mathrm{OIS}}(d,M)={\mathrm{OIS}}(d,M)  $.
If ${\left\{\rho_{m}^\text{in}\right\}_{m=1}^{M}}\in \underline{\mathrm{OIS}}(d,M)  $, all of them must be  pure states from the purity requirement.  This indicates that pure states may be better than mixed states in QPT for reducing the estimation error.
We provide the following two examples in Appendix \ref{appendixb} which belong to $ \underline{\mathrm{OIS}}(d,M) $ and achieve the lower bounds: SIC (symmetric informationally complete) states with the smallest $ M=d^2 $ and MUB (mutually unbiased bases) states for $ M=d(d+1) $. 
Moreover, we consider that   the system is restricted to a multi-qubit system where $d =2^m$ and each input state is an $m$-qubit tensor product state as  $ \rho_{j}=\rho^{(1)}_{j_{1}} \otimes \cdots \otimes \rho^{(m)}_{j_{m}} $ where $1 \leq j_{i} \leq M_{i} $ and there are $ M_i $ different types of single qubits for the $ i $-th qubit of the product states. Thus, the total number of these $ m $-qubit tensor product states is thus $M=\prod_{i=1}^{m}{M_i}$. We have the following theorem.
\begin{theorem}\label{theorem3}
	For $ m $-qubit product input states, we have $ M\operatorname{Tr}\big((V^{*} V^{T})^{-1}\big)\geq  20^{m}$ and $ \operatorname{cond}(V) \geq \sqrt{3^m}$.   These lower bounds are achieved simultaneously if and only if for each 
	$ 1\leq i \leq m $, $\left\{\rho^{(i)}_{j_i}\right\}_{i=1}^{M_i} \in \underline{\mathrm{OIS}}(2,M) $.
\end{theorem}
\begin{IEEEproof}
Assuming that the parameterization matrix in \eqref{probestate} of all the $i$-th single-qubit states $\left\{\rho_{j_i}^{(i)}\right\}_{j_i=1}^{M_i}$  is $V_i$. We thus have $V= V_1 \otimes V_2 \cdots \otimes V_m $. Therefore,
\begin{equation}
\begin{aligned}
M\operatorname{Tr}\left(\left(V^{*} V^{T}\right)^{-1}\right)&=M\prod_{i=1}^m \operatorname{Tr}\left(\left(V_i^{*} V_i^{T}\right)^{-1}\right) \geq M\prod_{i=1}^m \frac{20}{M_i}=20^{m},
\end{aligned}
\end{equation}	
and
\begin{equation}
\operatorname{cond}(V)=\prod_{i=1}^m \operatorname{cond}\left(V_i\right) \geq \prod_{i=1}^m \sqrt{3}=\sqrt{3^m}.
\end{equation}
The above two equalities hold if and only if for each 
$ 1\leq i \leq m $, $\left\{\rho^{(i)}_{j_i}\right\}_{i=1}^{M_i} \in \underline{\mathrm{OIS}}(2,M) $.
	\end{IEEEproof}

A similar proof for quantum detector tomography can also be found in \cite{xiao2021optimal}.
For one-qubit input states, they are in the Bloch sphere and  \cite{xiao2021optimal} has proved that platonic solids--tetrahedron, cube, octahedron, dodecahedron and icosahedron construct five examples, which belong to  $ \underline{\mathrm{OIS}}(2,M) $ with $ M=4, 6, 8, 12, 20 $ and  achieve one-qubit lower bounds. $ \underline{\mathrm{OIS}}(2,4) $ and $ \underline{\mathrm{OIS}}(2,6)$ are one-qubit SIC states and one-qubit MUB states, respectively.
For two-qubit product input states, Cube states (product states of two one-qubit MUB states) can achieve the lower bounds $ M\operatorname{Tr}\big((V^{*} V^{T})^{-1}\big)=400 $ and $ \operatorname{cond}(V)=3 $.
\subsection{Optimal measurement operators}
For our error bound  \eqref{finalall} or \eqref{tpbound}, the part depending on the measurement operators is $J\operatorname{Tr}\left(\left(C^{\dagger} C\right)^{-1}\right)$ and 
we need to minimize it. Since $\operatorname{cond}(I_{M} \otimes C)= \operatorname{cond}(I_M)\operatorname{cond}(C)=\operatorname{cond}(C) $, we also consider
the minimum  of $ \operatorname{cond}(C) $.

Assume $ P_{i,j} $ is the $ i $-th POVM element (i.e., a measurement operator) in the $ j $-th POVM  set where $1\leq j\leq J  $ and $ 1\leq i \leq n_j  $. Define the set size $ \mathcal{n}\triangleq[n_1, \cdots, n_J] $. Therefore, the total number of measurement operators $ L=\sum_{j=1}^{J} n_{j} $.  Let $\left\{\Omega_{i}\right\}_{i=1}^{d^{2}}$ be a complete set of $d$-dimensional traceless Hermitian matrices except $\Omega_{1}=I / \sqrt{d}$, and they satisfy $\operatorname{Tr}\left(\Omega_{i}^{\dagger} \Omega_{j}\right)=\delta_{i j}$
where $\delta_{i j}$ is the Kronecker function. Then we can parameterize the POVM element as
\begin{equation}
\label{para}
\begin{aligned}
P_{i,j} &=\sum_{a=1}^{d^{2}} \phi_{i,j}^{a} \Omega_{a}, \\
\end{aligned}
\end{equation}
where $\phi_{i,j}^{a} \triangleq \operatorname{Tr}\left(P_{i,j} \Omega_{a}\right)$ is real. Thus we use $ \phi_{i,j}\triangleq\left[\phi_{i,j}^{1}, \cdots, \phi_{i,j}^{d^2}\right]^{T} $ as the  parameterization of $P_{i,j}$ and the completeness constraint \eqref{povmcom} becomes
\begin{equation}\label{vecpovm}
\sum_{i=1}^{n_{j}} \phi_{i, j}=[\sqrt{d}, 0, \cdots, 0]^{T},
\end{equation}
and thus
\begin{equation}\label{vecmeasure2}
\sum_{i=1}^{n_{j}} \phi_{i, j}^{1}=\sqrt{d}, \quad\left\|\sum_{j=1}^{n_{j}} \phi_{i, j}\right\|=\sqrt{d}.
\end{equation}
Conversely, using \eqref{vecmeasure2}, we can also obtain \eqref{vecpovm}.
Define 
\begin{equation}
\tilde{C}=\left[\phi_{1,1}, \cdots \phi_{n_{1}, 1}, \phi_{1,2}, \cdots, \phi_{n_{J}, J}\right]^{T},
\end{equation}
which is the parameterization matrix of all the measurement operators in the basis of $\left\{\Omega_{i}\right\}_{i=1}^{d^{2}}$ and is a real matrix.  Since $\left\{\Omega_{i}\right\}_{i=1}^{d^{2}}$ and natural basis are both orthonormal basis, the relationship between $ C $ and $ \tilde{C} $ is $\tilde{C}=CU  $ where $ U $ is a unitary matrix.
Therefore, the eigenvalues of $ C^{\dagger}{C} $ and $\tilde C^{T}\tilde{C} $ are the same, and $ J\operatorname{Tr}\left(\left(C^{\dagger} C\right)^{-1}\right)= J\operatorname{Tr}\left(\left(\tilde C^{T} \tilde C\right)^{-1}\right)$, $ \operatorname{cond}(\tilde C) =\operatorname{cond}(C)$.

Similarly, 
we define the set of all $ d $-dimensional  measurement operators with set number $ J $ and set size $ \mathcal n $ as
\begin{equation}
\begin{aligned}
\mathcal M(d,J, \mathcal n) \triangleq\Big\{\left\{P_{i, j}\right\}_{i, j=1}^{n_{j}, J} \mid \forall 1 \leq j \leq J, \forall 1 \leq i \leq n_{j}, P_{i, j} \in \mathbb C^{d \times d}, P_{i, j}=P_{i, j}^{\dagger}, P_{i, j} \geq 0, 
\sum_{i=1}^{n_{j}} P_{i, j}=I_{d}, \Big\}.
\end{aligned}
\end{equation}
For general QPT, we have no  prior knowledge of the output states. From  \eqref{lsl}, the following conditions are equivalent: (i) the output states cannot be uniquely identified; (ii)  the measurement operators do not span the space of all $d$-dimensional Hermitian matrices; (iii) $C^{\dagger} C$ is singular; (iv)   $ J\operatorname{Tr}\left(\left(C^{\dagger} C\right)^{-1}\right) $ is infinite. From (i) and (iv), it is also reasonable to choose  $ J\operatorname{Tr}\left(\left(C^{\dagger} C\right)^{-1}\right) $ as the first cost function.
Thus we define the set of all the optimal  measurement operators to minimize  $J\operatorname{Tr}\left(\left(C^{\dagger} C\right)^{-1}\right)$ ($ \mathrm{OMO}_{1}(d,J, \mathcal n) $) as
\begin{equation}
\mathrm{OMO}_{1}(d,J, \mathcal n) \triangleq \underset{\left\{P_{i, j}\right\}_{i, j=1}^{n_{j}, J} \in \mathcal M(d,J, \mathcal n)}{\arg \min } J\operatorname{Tr}\left(\left(C^{\dagger} C\right)^{-1}\right).
\end{equation}
Since maximizing the robustness of the reconstructed output states with respect to measurement noise is equivalent  to minimizing the condition number, from \eqref{lsl}, we also choose $ \operatorname{cond}(C) $ as the second cost function. Thus, we define the set of all the optimal  measurement operators to minimize $ \operatorname{cond}(C) $ ($ \mathrm{OMO}_{2}(d,J, \mathcal n)  $)  as
\begin{equation}
\mathrm{OMO}_{2}(d,J, \mathcal n)  \triangleq \underset{\left\{P_{i, j}\right\}_{i, j=1}^{n_{j}, J} \in \mathcal M(d,J, \mathcal n)}{\arg \min } \operatorname{cond}(C).
\end{equation}

Then we define $ s\triangleq\sum_{j=1}^{J} \frac{d}{n_{j}} $ and two sets $ \underline{\mathrm{OMO}_{1}}(d,J, \mathcal n) $ and $ \underline{\mathrm{OMO}_{2}}(d,J, \mathcal n) $ to characterize the lower bounds of these two cost functions as 
\begin{equation}
\begin{aligned}
&\underline{\mathrm{OMO}_{1}}(d,J, \mathcal n) \triangleq \Big\{\left\{P_{i, j}\right\}_{i, j=1}^{n_{j}, J} \in \mathcal M(d,J, \mathcal n) \mid  J\operatorname{Tr}\left(\left(C^{\dagger} C\right)^{-1}\right)
=J\left(\frac{1}{s}+\frac{\left(d^{2}-1\right)^2}{J d-s}\right)
\Big\},
\end{aligned}
\end{equation}
\begin{equation}
\begin{aligned}
\underline{\mathrm{OMO}_{2}}(d,J, \mathcal n) \triangleq& \Big\{\left\{P_{i, j}\right\}_{i, j=1}^{n_{j}, J} \in \mathcal M(d,J, \mathcal n) \mid \operatorname{cond}({C})=\sqrt{\frac{\left(d^{2}-1\right)s}{J d-s}}\Big\},
\end{aligned}
\end{equation}
where the proof is given as in Theorem \ref{theorem2m}. 
We also define the set of all the optimal measurement operators ($ \mathrm{OMO}(d,J, \mathcal n)  $) as
\begin{equation}
\begin{aligned}
\mathrm{OMO}(d,J, \mathcal n) \triangleq \mathrm{OMO}_{1}(d,J, \mathcal n)  \cap \mathrm{OMO}_{2}(d,J, \mathcal n),
\end{aligned}
\end{equation}
and the set of all the optimal measurement operators achieving both of the lower bounds ($ \underline{\mathrm{OMO}}(d,J, \mathcal n) $)  as 
\begin{equation}
\begin{aligned}
\underline{\mathrm{OMO}}(d,J, \mathcal n)\triangleq \underline{\mathrm{OMO}_{1}}(d,J, \mathcal n) \cap \underline{\mathrm{OMO}_{2}}(d,J, \mathcal n).
\end{aligned}
\end{equation}
Thus, $ \mathrm{OMO}(d,J, \mathcal n) $, when it is non-empty, should minimize  $J\operatorname{Tr}\left(\left(C^{\dagger} C\right)^{-1}\right)$ and minimize $\operatorname{cond}(C)$ \emph{simultaneously}. When $ \underline{\mathrm{OMO}}(d,J, \mathcal n) $ is non-empty, it can  achieve the lower bounds of  $J\operatorname{Tr}\left(\left(C^{\dagger} C\right)^{-1}\right)$ and $\operatorname{cond}(C)$ \emph{simultaneously}.
Then we present the following theorem for $\underline{\mathrm{OMO}}(d,J, \mathcal n)  $.
\begin{theorem}
	\label{theorem2m}
	For a $ d $-dimensional quantum process, let the total number of POVM sets be $ J $ and the number of POVM elements in the $ j $-th POVM set be $ n_j $. Then $$  J\operatorname{Tr}\left(\left(C^{\dagger} C\right)^{-1}\right) \geq J\left(\frac{1}{s}+\frac{\left(d^{2}-1\right)^{2}}{J d-s}\right) $$ and $$\operatorname{cond}(C)\geq \sqrt{\frac{\left(d^{2}-1\right) s}{J d-s}}. $$ These lower bounds are achieved simultaneously; i.e., $ \left\{P_{i, j}\right\}_{i, j=1}^{n_{j}, J}\in \underline{\mathrm{OMO}}(d,J, \mathcal n)  $, if and only if the eigenvalues of $ C^{\dagger} C$     are  $ v_{1}=s $ and $v_{2}=\dots=v_{d^2}=\frac{Jd-s}{d^2-1} $.
	
\end{theorem}
\begin{IEEEproof}
	Denote  the eigenvalues of $ C^{\dagger}{C} $ or $\tilde C^{T}\tilde{C} $ as $  v_{1} \geq v_{2} \geq \cdots \geq v_{d^2} >0 $. Since all the measurement operators are positive semidefinite, we have
	\begin{equation}
	\operatorname{Tr}\left((P_{m, j})^{\dagger} P_{n, j}\right)=\left(\phi_{m, j}\right)^{T} \phi_{n, j} \geq 0.
	\end{equation}
	Using \eqref{vecpovm}, we have 
	\begin{equation}
	\begin{aligned}
	d&=\left\|\sum_{i=1}^{n_{j}} \phi_{i, j}\right\|^{2}=\sum_{i=1}^{n_{j}}\left\|\phi_{i, j}\right\|^{2}+\sum_{m \neq n}\left(\phi_{m, j}\right)^{T} \phi_{n, j} \\
	&\geq \sum_{i=1}^{n_{j}}\left\|\phi_{i, j}\right\|^{2}.
	\end{aligned}
	\end{equation}
	Therefore,
	\begin{equation}
	\sum_{i=1}^{d^{2}} v_{i}=\operatorname{Tr}\left(\tilde{C}^{T} \tilde C\right)=\sum_{j=1}^{J}\left(\sum_{i=1}^{n_{j}}\left\|\phi_{i, j}\right\|^{2}\right) \leq J d.
	\end{equation}	
	Since
	\begin{equation}
	\sum_{i=1}^{n_{j}}\left(\phi_{i, j}^{1}\right)^{2} \geq \frac{\left(\sum_{i=1}^{n_{j}} \phi_{i, j}^{1}\right)^{2}}{n_{j}}=\frac{d}{n_{j}},
	\end{equation}	
	the first diagonal element of $\tilde{C}^{T} \tilde C$  is 
	\begin{equation}
	\begin{aligned}
	\left(\tilde{C}^{T} \tilde C\right)_{11}&=\sum_{j=1}^{J}\left(\sum_{i=1}^{n_{j}}\left(\phi_{i, j}^{1}\right)^{2}\right) \geq \sum_{j=1}^{J} \frac{d}{n_{j}},
	\end{aligned}
	\end{equation}
	and thus 
	\begin{equation}
	v_{1}\geq \left(\tilde{C}^{T} \tilde C\right)_{11} \geq \sum_{j=1}^{J} \frac{d}{n_{j}}.
	\end{equation}
	To minimize $J\operatorname{Tr}\left(\left(C^{\dagger} C\right)^{-1}\right)$ and $\operatorname{cond}(C)$, the optimization problems can be relaxed 
	as  follows	
	\begin{equation}
	\label{umsem}
	\begin{aligned}
	\text { min } & \sum_{i=1}^{d^2} \frac{J}{v_{i}} \\
	\text { s.t. } & \sum_{i=1}^{d^2} v_{i}\leq Jd, v_{1}\geq\sum_{j=1}^{J} \frac{d}{n_{j}},
	\end{aligned}
	\end{equation}
	and 
	\begin{equation}
	\label{robustm}
	\begin{aligned}
	\text { min } & \sqrt{ \frac{v_{1}}{v_{d^2}}} \\
	\text { s.t. } & \sum_{i=1}^{d^2} v_{i}\leq Jd,v_{1}\geq\sum_{j=1}^{J} \frac{d}{n_{j}}.
	\end{aligned}
	\end{equation}
 Using the Lagrange multiplier method,  we have
	\begin{equation}
	\begin{aligned}
	&J\operatorname{Tr}\left(\left(C^{\dagger} C\right)^{-1}\right)=	J\operatorname{Tr}\left(\left(\tilde{C}^{T} \tilde C\right)^{-1}\right)\geq J\left(\frac{1}{s}+\frac{\left(d^{2}-1\right)^2}{J d-s}\right),
	\end{aligned}
	\end{equation}
	and
	\begin{equation}
	\operatorname{cond}({C})\geq \sqrt{\frac{s}{(Jd-s)/(d^2-1)}}=\sqrt{\frac{\left(d^{2}-1\right)s}{J d-s}}.
	\end{equation}
	These lower bounds	are achieved simultaneously if and only if $ v_{1}= \sum_{j=1}^{J} \frac{d}{n_{j}} $ and $v_{2}=\dots=v_{d^2}=\frac{Jd-v_1}{d^2-1}  $.
\end{IEEEproof}

Since $ \mathcal M(d,J, \mathcal n) $ is a closed space, there always exist non-empty $ \mathrm{OMO}_{1}(d,J, \mathcal n) $ and $ \mathrm{OMO}_{2}(d,J, \mathcal n) $ for  arbitrary set number $ J $ and set size $  \mathcal n $. However, it is not clear whether this also holds for $ \mathrm{OMO}(d,J, \mathcal n) $. For the lower bounds of these two cost functions, if one of them is achieved, we have $ v_{1}=\sum_{j=1}^{J} \frac{d}{n_{j}}, v_{2}=\dots=v_{d^2}=\frac{Jd-v_1}{d^2-1}$ and thus 
\begin{equation}\label{omo}
\begin{aligned}
\underline{\mathrm{OMO}}(d,J, \mathcal n)= \underline{\mathrm{OMO}_{1}}(d,J, \mathcal n) = \underline{\mathrm{OMO}_{2}}(d,J, \mathcal n).
\end{aligned}
\end{equation}
If the two lower bounds can not be achieved, the above three sets are empty.
Therefore, \eqref{omo} always holds for any $J$ and $\mathcal n $.
Our results show that if $ \left\{P_{i, j}\right\}_{i, j=1}^{n_{j}, J} \in  \underline{\mathrm{OMO}}(d,J, \mathcal n) $, then in each POVM set, these POVM elements are orthogonal; i.e., $ \operatorname{Tr}\left((P_{m, j})^{\dagger} P_{n, j}\right)=0$ $(m\neq n) $ from the first constraint. To achieve these lower bounds, we must have $ J\geq d $ because the maximum number of  orthogonal POVM elements in one set is $ d $. It is an open problem as to whether  $ \underline{\mathrm{OMO}}(d,J, \mathcal n) $ is non-empty for arbitrary set number $ J $ and set sizes $ n_j $ ($ 1\leq j\leq J $). When it is non-empty, we have $\underline{\mathrm{OMO}}(d,J, \mathcal n)={\mathrm{OMO}}(d,J, \mathcal n)  $.

Here, we give one example of $ \underline{\mathrm{OMO}}(d,J, \mathcal n) $: MUB (mutually unbiased bases) measurement with $ J=d+1$ and $ n_j=d $ for $ 1\leq j \leq d+1 $, which can achieve these lower bounds.
Two sets of orthogonal
bases $ \mathcal{H}^{m}=\{\left|\psi_{i}^{m}\right\rangle \}_{i=1}^{d}$ and $ \mathcal{H}^{n}=\{\left|\psi_{j}^{n}\right\rangle\}_{j=1}^{d} $ are called
mutually unbiased if and only if \cite{PhysRevLett.105.030406}
\begin{equation}
\label{mub}
\left|\left\langle\psi_{i}^{m} | \psi_{j}^{n}\right\rangle\right|^{2}=\left\{\begin{array}{ll}
\frac{1}{d} & \text { for } m \neq n, \\
\delta_{i j} & \text { for } m=n.
\end{array}\right.
\end{equation}
For a prime $p$ and a positive integer $k$, there exist maximally  $d+1$ sets of mutually unbiased bases in Hilbert spaces of prime-power dimension $d=p^{k}$ \cite{mubreview}. For other values, it is still an open problem as to whether there exist $d+1$ sets of mutually unbiased bases. The corresponding MUB measurements are   $\left\{\left|\psi_{i}^{m}\right\rangle\left\langle\psi_{i}^{m}\right|\right\}_{m, i=1}^{d+1, d}$.
Then we give the following proposition with a similar proof to that of Proposition 2 in \cite{xiao2021optimal}.
\begin{proposition}
	\label{lem22}
	$ d $-dimensional 	MUB measurements (when they exist) belong to $\underline{\mathrm{OMO}}(d,J, \mathcal n) $ with $ J=d+1 $ and $ n_j=d $ for $ 1\leq j \leq d+1 $.
\end{proposition}
\begin{IEEEproof}
	We assume that the standard singular value decomposition (SVD) of $C$ is $C=U_{c} \Sigma V^{\dagger}_{c}$. Thus, using \eqref{aa} and \eqref{mub}, we have
	\begin{equation}
	C C^{\dagger}=\left[\begin{array}{cccc}
I_d & \left(\frac{1}{d}\right)_d & \cdots & \left(\frac{1}{d}\right)_d \\
\left(\frac{1}{d}\right)_d & I_d & \cdots & \left(\frac{1}{d}\right)_d \\
\vdots & \vdots & \vdots & \vdots \\
\left(\frac{1}{d}\right)_d & \left(\frac{1}{d}\right)_d & \cdots & I_d
\end{array}\right]=U_{c} \Sigma \Sigma^T U^{\dagger}_{c}
	\end{equation}
	where $\left(\frac{1}{d}\right)_d$ denotes the $d \times d$ matrix where all the elements are $\frac{1}{d}$. Therefore,
	\begin{equation}
		\begin{aligned}
	C^{\dagger}C & =V_{c} \Sigma^T U^T_{c} U_{c} \Sigma V^{\dagger}_{c} =V_{c} \operatorname{diag}\left(d+1, I_{d^2-1}\right) V^{\dagger}_{c}.
	\end{aligned}
	\end{equation}
Thus the eigenvalues of $ C^{\dagger} C $ are $ v_1=d+1, v_2=\cdots=v_{d^2}=1 $ for MUB measurements and we can then calculate $ J\operatorname{Tr}\left(\left(C^{\dagger} C\right)^{-1}\right) = d^3+d^2-d $ and $ \operatorname{cond}(C)=\sqrt{d+1} $, which achieve the lower bounds.
	\end{IEEEproof}

\begin{remark}
	SIC-POVM are usually thought as optimal (in certain senses) measurements in quantum physics. The simplest mathematical  definition of  SIC-POVM is a set of $d^{2}$ normalized vectors $\left|\psi_{k}\right\rangle$ in $\mathbb{C}^{d}$ satisfying \cite{sic}
	\begin{equation}
	\label{sic2}
	\left|\left\langle\psi_{j} | \psi_{k}\right\rangle\right|^{2}=\frac{1}{d+1}, j \neq k.
	\end{equation}
	The corresponding POVM elements are $P_{i}=\frac{1}{d}\left|\psi_{i}\right\rangle\left\langle\psi_{i}\right|$ for $ 1\leq i \leq d^2 $ and $ \sum_{i=1}^{d^2}P_i=I $ with $ J=1 $, $ n_1=d^2 $.
	However, here SIC-POVM does not belong to $\underline{\mathrm{OMO}}(d,J, \mathcal n)$ because these POVM elements are not orthogonal with each other. Therefore, the first constraint is not satisfied. 	We can  calculate $ J\operatorname{Tr}\left(\left(C^{\dagger} C\right)^{-1}\right) = d^3+d^2-d $ and $ \operatorname{cond}(C)=\sqrt{d+1} $ for SIC-POVM. These values are the same as MUB measurement which has $ J=d+1 $ and $ n_j=d $ for $ 1\leq j \leq d+1 $. Thus we conjecture that SIC-POVM belongs to $\mathrm{OMO}(d,J, \mathcal n)$ for $ J=1 $, $ n_1=d^2 $.
\end{remark}
\begin{remark}
Our analysis for optimal measurement operators can also be applied to QST.
A similar problem for QST has been discussed in \cite{Qi2013}. However, their work is restricted to optimal projective measurement, while our result considers general POVM. Miranowicz \emph{et al.}  \cite{PhysRevA.90.062123} also considered the minimum condition number  for the optimal  measurement operators in QST and proposed generalized Pauli operators whose condition number is $ 1 $. However, generalized Pauli operators are realized from projective measurements and their condition number  does not directly link to the experimental data.
\end{remark}

With  input states like SIC states or MUB states belonging to $ \underline{\mathrm{OIS}}(d,M) $ and measurement operators like MUB measurements belonging to $ \underline{\mathrm{OMO}}(d,J, \mathcal n)$, the error upper bound of \eqref{finalall} becomes $ O\left(\frac{d^{4}\operatorname{Tr}(F)}{\sqrt{N}}\right) $.
The limitation of our error upper bound is that it might be loose with respect to the system dimension $ d $ and we leave it an open problem to characterize $ d $ more accurately. We can obtain a tighter upper bound if  we use  the natural basis states $\{|j\rangle\langle k|\}_{1\leq j,k\leq d}$ and $ M=d^2 $ with MUB measurements. Note that in experiments, all of the quantum states are  Hermitian. Hence, we cannot actually use $|j\rangle\langle k|$ ($j\neq k$) as input quantum states. According to \cite{qci}, since we have
\begin{equation}\label{jkstate}
\begin{aligned}
|j\rangle\langle k|=&|+\rangle\langle+|+\mathrm i|-\rangle\langle-|-\frac{1+\mathrm i}{2}|j\rangle\langle j|-\frac{1+\mathrm i}{2}| k\rangle\langle k|,
\end{aligned}
\end{equation}
where $|\pm\rangle=(|j\rangle\pm|k\rangle)/\sqrt2$,
$\mathcal{E}(|j\rangle\langle k|)$ can be obtained as
\begin{equation}\label{eq10}
\begin{array}{rl}
\mathcal{E}\left(|j\rangle\langle k|\right)=&\mathcal{E}(|+\rangle\langle +|)+\mathrm i\mathcal{E}(|-\rangle\langle -|) -\dfrac{1+\mathrm i}{2}\mathcal{E}(|j\rangle\langle j|)-\dfrac{1+\mathrm i}{2}\mathcal{E}(|k\rangle\langle k|).\\
\end{array}
\end{equation}
Using natural basis states which we define as all the states needed by the RHS of \eqref{jkstate}, Wang \emph{et al.} \cite{8022944} have proved that $ B $ is a permutation matrix. Thus, the error analysis in Step 2 becomes 
\begin{equation}
\begin{aligned}
\|\hat{D}-D\|&=\left\|\operatorname{vec}^{-1}\left(B^{\dagger} \operatorname{vec}(\hat{A})\right)-\operatorname{vec}^{-1}\left(B^{\dagger} \operatorname{vec}(A)\right)\right\|=\left\|B^{\dagger}(\operatorname{vec}(\hat{A})-\operatorname{vec}(A))\right\|=\|\hat{A}-A\|,
\end{aligned}
\end{equation}
and the error analysis in other steps are the same as our analysis. Therefore, we obtain a tighter upper bound as $O\left(\frac{d^{3}\operatorname{Tr}(F)}{\sqrt{N}}\right)$. 
%\vspace*{-15pt}
\subsection{On the different types of input quantum states and measurement operators}
In practice,  the different types of input quantum states $ M $ and measurement operators $ J $ may vary and here we discuss the MSE scaling with respect to $ M $ and $ J $.
We first allow $M$ to change, with the dimension $d$, the measurement operators and the copy number of each input state $N$ given. The different input states are randomly generated as \cite{MISZCZAK2012118,qetlab} or as  \cite{wang2019twostage} for truncated coherent states. Coherent states are more straightforward to be prepared than Fock states in quantum optics  and have been used in QPT in \cite{science1162086,rahimi2011quantum}.
Each scenario with $ N $ copies generates the input states $ \rho_{m}^\text{in} $ according to a certain probability distribution and the total copies are $ N_t=NM $, based on which $ \mathrm{E}(\cdot) $ denotes the expectation, different from $ \mathbb E(\cdot) $ in Theorem \ref{mainthe}.
Define  $\mathrm f_{m}\triangleq\operatorname{vec}\left(\rho_{m}^{\text{in}}\right)-\mathrm{E}\left(\operatorname{vec}\left(\rho_{m}^{\text{in}}\right)\right)$, which is i.i.d. with respect to the subscript $i$. Thus we have
\begin{align*}
&\quad\mathrm{E}\left\{ \operatorname{Tr}\left(\left(V^{*} V^{T}\right)^{-1}\right)\right\} \\
&=\frac{1}{M} \operatorname{Tr}\Big\{M \mathrm{E}\Big[\Big(\sum_{m=1}^{M}\mathrm{E}\left(\operatorname{vec}\left(\rho_{m}^{\text{in}}\right)^{*}\right)+\mathrm f_{m}^{*}\Big)\cdot\Big(\mathrm{E}\Big(\operatorname{vec}\left(\rho_{m}^{\text{in}}\right)^{T}\Big)+\mathrm f_{m}^{T}\Big)\Big)^{-1}\Big]\Big\} \\
&=\frac{1}{M} \operatorname{Tr}\Big\{M \mathrm{E}\Big[\Big(M \mathrm{E}\left(\operatorname{vec}\left(\rho_{m}^{\text{in}}\right)^{*}\right) \mathrm{E}\left(\operatorname{vec}\left(\rho_{m}^{\text{in}}\right)^{T}\right)+\sum_{m=1}^{M} \mathrm f_{m}^{*}\mathrm  f_{m}^{T}\Big)^{-1}\Big]\Big\} \\
&=\frac{1}{M} \operatorname{Tr}\Big\{\Big[\mathrm{E}\left(\operatorname{vec}\left(\rho_{m}^{\text{in}}\right)^{*}\right) \mathrm{E}\Big(\operatorname{vec}\left(\rho_{m}^{\text{in}}\right)^{T}\Big)+\mathrm{E}\Big(\frac{1}{M} \sum_{m=1}^{M} \mathrm f_{m}^{*}\mathrm  f_{m}^{T}\Big)\Big]^{-1}\Big\}. \numberthis \label{rl}
\end{align*}
According to the central limit theorem, as $M$ tends to infinity, $\mathrm{E}(\frac{1}{M}\sum_{m=1}^M \mathrm f_m^{*}\mathrm f_m^T)$ converges to a fixed matrix \cite{wang2019twostage}. Therefore the expectation of $ \operatorname{Tr}\left(\left(V^{*} V^{T}\right)^{-1}\right) $ is
\begin{equation}\label{vm}
\begin{aligned}
&\mathrm{E}\left\{ \operatorname{Tr}\left(\left(V^{*} V^{T}\right)^{-1}\right)\right\} = O\left(\frac{1}{{M}}\right). \\
\end{aligned}
\end{equation}
In Step 1, since $ N $ is given, using \eqref{qstmse}, we have
\begin{equation}\label{smalllam}
\begin{aligned}
\mathrm E\left\|\operatorname{col}_{m}(\hat A^{T})-\operatorname{col}_{m}( A^{T})\right\|^{2}=O\left(1\right).
\end{aligned}
\end{equation}
In Step 2, assume that the SVD of $ \left(V^{*} V^{T}\right)^{-1} V^{*} $ is
\begin{equation}
\left(V^{*} V^{T}\right)^{-1} V^{*}=U\left[
\Sigma_{1}, 0
\right] W^{\dagger},
\end{equation}
where $  \Sigma_{1}$ is a $ d^2\times d^2 $ diagonal matrix, and $ U $ and $ W $ are unitary. Using \eqref{vv} and \eqref{vm}, $ \mathrm E\left\|\Sigma_{1}\right\|^{2} = O\left(\frac{1}{{M}}\right) $. 
Let 
\begin{equation}
h_m=W^{\dagger}\left(\operatorname{col}_{m}(\hat A)-\operatorname{col}_{m}( A)\right)=\left[\begin{array}{l}
h_m^{(1)} \\
h_m^{(2)}
\end{array}\right],
\end{equation}
where $ h_m^{(1)} $ is a $ d^2\times 1 $ vector, $ h_m^{(2)} $ is a $ (M-d^2)\times 1 $ vector and each element in $ h_m $ scales as $ O(1) $ from \eqref{smalllam}. Then for the error in Step 2, we have
\begin{align*}
&\quad\mathrm{E}\left\|R^{T}\left(I_{d^2} \otimes\left(V^{*} V^{T}\right)^{-1} V^{*}\right)(\operatorname{vec}(\hat{A})-\operatorname{vec}(A))\right\|^{2} \\
&=\mathrm{E}\sum_{m=1}^{d^{2}}\left\|\left(V^{*} V^{T}\right)^{-1} V^{*}\left(\operatorname{col}_{m}(\hat A)-\operatorname{col}_{m}( A)\right)\right\|^{2} \\
&=\mathrm{E}\sum_{m=1}^{d^{2}}\left\|U\left[\Sigma_{1}, 0\right]\left[\begin{array}{l}
h_{m}^{(1)} \\
h_{m}^{(2)}
\end{array}\right]\right\|^{2} \\
&=\mathrm{E}\sum_{m=1}^{d^{2}}\left\|\Sigma_{1} h_{m}^{(1)}\right\|^{2} =O\left(\frac{1}{{M}}\right). \numberthis
\end{align*}
Since Steps 3--4 are not related to the scaling on $ M $,
the final error scaling on $ M $ is
\begin{equation}\label{statem}
\mathrm{E}\|\hat{X}-X\|=O\left(\frac{1}{\sqrt{M}}\right).
\end{equation}
From our error upper bound in \eqref{finalall}, using \eqref{vm}, we have
\begin{equation}
\begin{aligned}
&\mathrm{E}\left\{ M\operatorname{Tr}\left(\left(V^{*} V^{T}\right)^{-1}\right)\right\} = O\left(1\right), \\
\end{aligned}
\end{equation}	
and thus the scaling  of \eqref{finalall} is $O\left(1\right) $ with given $ N $ and $ d $. But the accurate scaling on $ M $ is $ \mathrm{E}\|\hat{X}-X\|=O\left(\frac{1}{\sqrt{M}}\right) $.
Therefore, our error upper bound \eqref{finalall} is not always tight with respect to $ M $.

Then we consider the MSE scaling of  our TSS on $ J $. With given input states, $ N $ and $ d $, we can also generate  different measurement operators 
according to certain probability distributions.
Similarly, we can prove 
\begin{equation}
\mathrm E\left\|\operatorname{col}_{m}(\hat A^{T})-\operatorname{col}_{m}( A^{T})\right\|^{2}=O(1),
\end{equation}
and
\begin{equation}
\begin{aligned}
&\mathrm{E}\left\{ J\operatorname{Tr}\left(\left(C^{\dagger} C\right)^{-1}\right)\right\} = O\left(1\right). \\
\end{aligned}
\end{equation}
Therefore, our error upper bound \eqref{finalall} is tight with respect to $ J$.

\section{Numerical examples}\label{numerical}
To perform measurement on the output states,
there are  different measurement bases such as SIC-POVM \cite{sic}, MUB measurement \cite{mubreview}, and Cube bases \cite{PhysRevA.78.052122}. In this section, we use Cube bases, as it is relatively easy to be implemented in experiments. For one-qubit systems, the Cube bases are $\left\{\frac{I \pm \sigma_{x}}{2}, \frac{I \pm \sigma_{y}}{2}, \frac{I \pm \sigma_{z}}{2}\right\}$ where $ \sigma_{x} $, $ \sigma_{y} $, $\sigma_{z}$ are Pauli matrices. For  multi-qubit systems, the Cube bases are the tensor products of one-qubit Cube bases.
\subsection{Performance illustration}
We consider a $ 4 $-dimensional quantum system and the TP process matrix $ X $ is determined by $\{\mathcal A_i\}_{i=1}^{3}$ where
\begin{equation}\label{4rand}
\begin{aligned}
&\mathcal A_{1}=U_{1} \operatorname{diag}(0.5,0.4,0,0), \\
&\mathcal A_{2}=U_{2} \operatorname{diag}(0.1,0.2,0,0), \\
&\mathcal A_{3}=U_{3} \sqrt{I-\mathcal A_{1}^{\dagger} \mathcal A_{1}-\mathcal A_{2}^{\dagger} \mathcal A_{2}},
\end{aligned}
\end{equation}
where $ U_1 $, $ U_2 $ and $ U_3 $ are random unitary matrices \cite{MISZCZAK2012118,qetlab}.

\begin{figure}
	\centering
	\subfigure{
		\begin{minipage}[b]{1\textwidth}
		\centering	\includegraphics[width=4.5in]{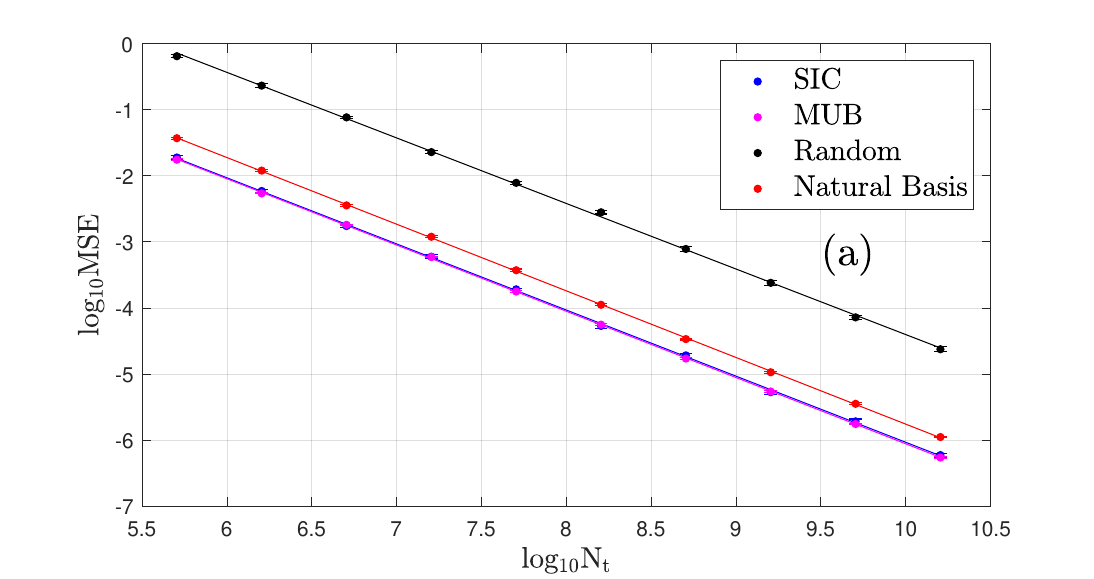}
		\end{minipage}
	}
	\subfigure{
		\begin{minipage}[b]{1\textwidth}
		\centering	\includegraphics[width=4.5in]{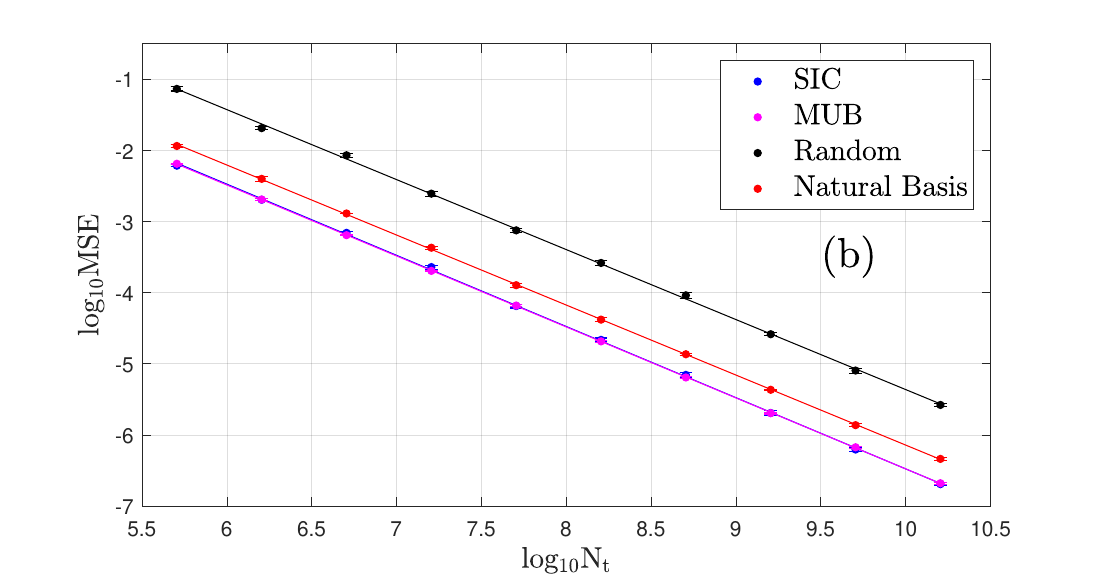}
		\end{minipage}
	}
	\caption{Log-log plot of MSE versus  the total number of copies $N_t$  using SIC states, MUB states, random states and natural basis states. (a) TP process; (b) non-TP process.} \label{mse}
\end{figure}

We use four different sets of probe states.
The first set is SIC state which is given in \eqref{sic} in Appendix \ref{appendixb}  and $ M=16 $. The second set is MUB states which is given in \eqref{mubbase} in Appendix \ref{appendixb} and  $ M=20 $. The third one is random states where we randomly generate $ 20 $ input states using the algorithm in \cite{MISZCZAK2012118,qetlab}. The fourth set is natural basis states in the RHS of \eqref{jkstate}.
The result of MSE (mean squared error) versus the total number of copies $ N_t=MN $ is shown in Fig. \ref{mse}(a), where $ \operatorname{MSE}= \mathbb E||\hat X-X||^2 $ and we make all of the four sets have the same $ N_t $. For each number of copies $N_t$, we repeat our algorithm $ 20 $ times and obtain the average MSE and error bars. 
For all the four sets of input states,
the scaling of MSE is basically $ O(1/N_t)$ which satisfies Theorem \ref{mainthe}. Moreover, the MSEs of SIC states, MUB states, natural basis states are all smaller than random states. The MSEs of SIC states and MUB states are quite similar and both smaller than natural basis states.
For non-TP processes, 
we consider the same $ \mathcal A_{1}, \mathcal A_{2} $ and different $ \mathcal A_{3} $ as
\begin{equation}
\mathcal A_{3}=U_{3} \sqrt{U_4\operatorname{diag}(1,0.8,0.7,0.5)U_4^{\dagger}-\mathcal A_{1}^{\dagger} \mathcal A_{1}-\mathcal A_{2}^{\dagger}\mathcal A_{2}},
\end{equation}
where $ U_4 $ is a random unitary matrix.
The result for non-TP processes is shown in Fig. \ref{mse}(b) which is similar to Fig. \ref{mse}(a).
The MSE of non-TP processes is a little smaller than that of TP processes because the constraint for TP processes is stricter than that for non-TP processes.

\begin{figure}
	\centering
	\includegraphics[width=4.5in]{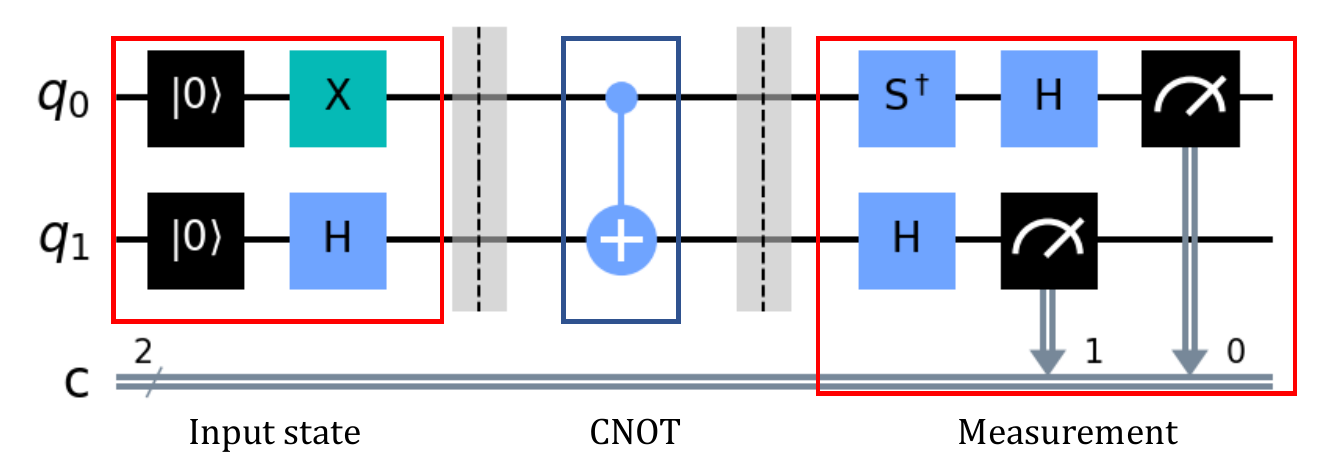}
	\centering{\caption{An example of the IBM Composer. The input state is $  \frac{I-\sigma_{z}}{2}\otimes  \frac{I + \sigma_{x}}{2} $ and the measurement operators are $ \frac{I \pm \sigma_{y}}{2} \otimes \frac{I \pm \sigma_{x}}{2}  $.}\label{cnot}}
\end{figure}

We also test our algorithm on the IBM Quantum device \cite{ibm}. Here we perform QPT on the CNOT gate defined as
\begin{equation}
\text{CNOT}=\left[\begin{array}{llll}
1 & 0 & 0 & 0 \\
0 & 0 & 0 & 1 \\
0 & 0 & 1 & 0 \\
0 & 1 & 0 & 0
\end{array}\right],
\end{equation}
in the IBM Quantum Composer. For the input states,
we use states  $\left\{\frac{I + \sigma_{x}}{2}, \frac{I + \sigma_{y}}{2}, \frac{I \pm \sigma_{z}}{2}\right\}^{\otimes 2}$ which can be generated by applying Hadamard and $S^\dagger$ gates to qubits initialized in the $|0\rangle$ state \cite{dang2021process}. In addition, we use Cube measurements. Fig. \ref{cnot} is an example of IBM Quantum Composer where the input state is $  \frac{I-\sigma_{z}}{2}\otimes  \frac{I + \sigma_{x}}{2} $ and the measurement operators are $ \frac{I \pm \sigma_{y}}{2} \otimes \frac{I \pm \sigma_{x}}{2}  $.

\begin{figure}
	\centering
	\includegraphics[width=4.5in]{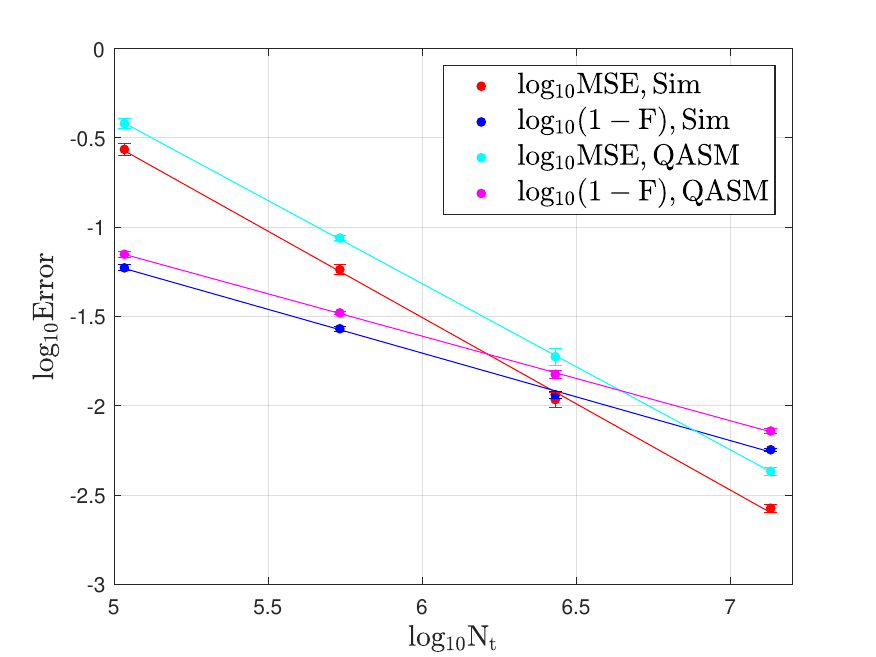}
	{\caption{Log-log plot of MSE and infidelity versus  the total number of copies $N_t$ for the CNOT process using simulation (Sim) by MATLAB and $ibmq\_qasm\_simulator$ (QASM).}\label{cnotmse}}
\end{figure}

\begin{figure}
	\centering
	\includegraphics[width=4.5in]{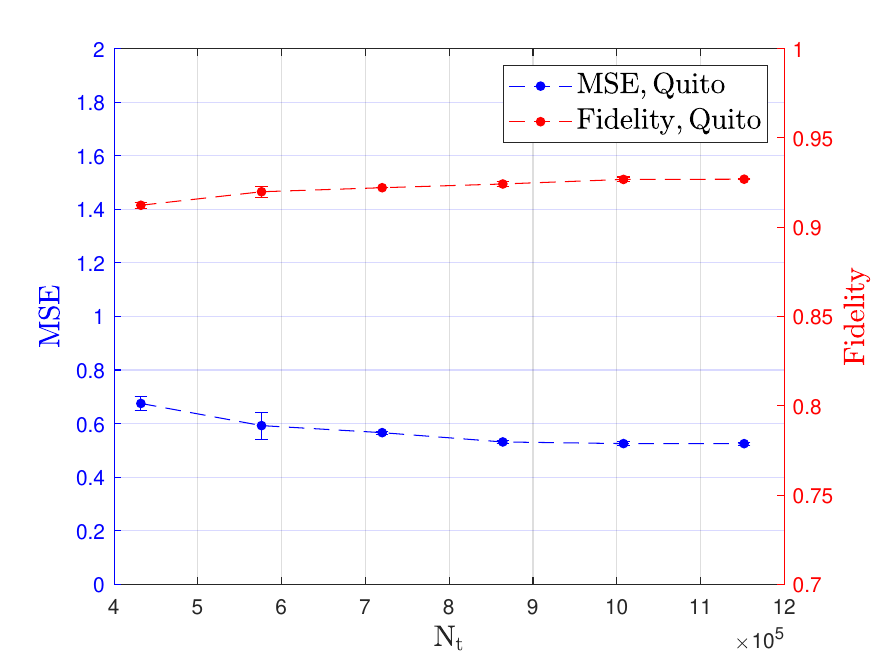}
	\centering{\caption{MSE and infidelity versus  the total  number of copies $N_t$ for the CNOT process using $ibmq\_quito$ 5-qubit system.}\label{quitomse}}
\end{figure}

If we have prior knowledge that the process is unitary, some references have proposed effective identification algorithms \cite{Gutoski2014,WANG2019269}.
Here we assume that we do not have prior knowledge that the unknown process is in essence unitary. 
We perform these numerical examples using the MATLAB on a classical computer and the $ibmq\_qasm\_simulator$. The  total numbers of copies are $N_t=1.08\times10^5, 5.40\times10^5, 2.70\times10^6, 1.35\times10^7 $ and the experiments are repeated 5 times. Then we apply our TSS algorithm and the results are shown in Fig. \ref{cnotmse}. Besides MSE, we also consider another common fidelity metric, defined as \cite{PhysRevA.82.042307}
\begin{equation}
F(\hat{X},X)\triangleq \left[\operatorname{Tr} \sqrt{\sqrt{\hat{X}} X\sqrt{\hat{X}}}\right]^{2}/\left[\operatorname{Tr}\left(X\right) \operatorname{Tr}\left(\hat{X}\right)\right],
\end{equation}
and the correspond infidelity is defined as $1-F(\hat{X},X)$.
From Fig. \ref{cnotmse}, the scalings of the MSEs from $ibmq\_qasm\_simulator$ and simulation are also both $ O(1/N_t)$ which satisfies Theorem \ref{mainthe}. The scalings of the infidelities are both  $ O(1/\sqrt{N_t})$ because the CNOT process is rank-deficient. A similar infidelity scaling has also been studied in QST \cite{PhysRevLett.105.030406,Qi2017}. In addition, the errors between simulation and $ibmq\_qasm\_simulator$ are close. Then we apply our algorithm on
the $ibmq\_quito$ 5-qubit system. The MSE  and fidelity are presented in Fig. \ref{quitomse} where  the  total number of copies are $N_t=4.32\times10^5, 5.76\times10^5, 7.20\times10^5, 8.64\times10^5, 10.08\times10^5, 11.52\times10^5 $, respectively, and we repeat $ 3 $ times.
These results are both worse than $ibmq\_qasm\_simulator$ and simulation because there exist state preparation error and readout error in $ibmq\_quito$ \cite{ibm}. The average CNOT error is $ 1.253\times10^{-2} $ and average readout error is $ 4.430\times10^{-2} $ for $ibmq\_quito$. Thus, the approximate fidelity is $ (1- 1.253\times10^{-2})(1-4.430\times10^{-2})=0.9437$.

\subsection{Performance comparison}
Here, we compare the performance between our TSS algorithm  and the convex optimization method in \cite{unitary}. Baldwin \emph{et al.} \cite{unitary} formulated the TP QPT problem as a convex optimization problem  
\begin{equation}
\begin{aligned}
\min & \sum_{m, l}\left|\mathcal g_{m l}-\operatorname{Tr}\left(\mathcal O_{m l}^{\dagger} \hat X\right)\right|^{2} \\
\text { s.t. } & \sum_{j,k } \hat X_{j k} E_{k}^{\dagger} E_{j}=I, \\
& \hat X=\hat X^{\dagger}, \quad \hat X \geq 0,
\end{aligned}
\end{equation}
where $\mathcal g_{m l}$ is measurement data, $ \mathcal O_{m l} $ is a constant matrix which is the tensor product between the transpose of the $ m $-th input states and $ l $-th measurement bases where $1\leq m\leq M$, $1\leq l\leq L   $, and $ \hat X $ is the estimated process matrix. The detailed description can be found in equation (13) in \cite{unitary}.  This problem can be solved as a convex
semidefinite program, which is realized by CVX \cite{cvx,gb08}.

\begin{figure}
	\centering
	\includegraphics[width=4.5in]{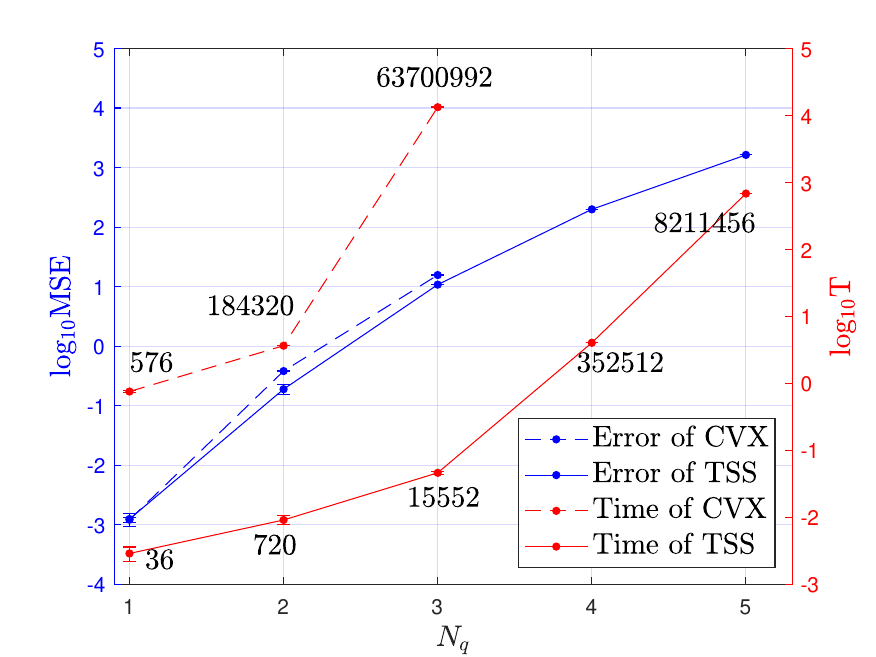}
	\centering{\caption{Using random quantum states, running time $T$ and the MSE versus qubit number $N_q$ for the convex optimization method in \cite{unitary} and our TSS algorithm. The corresponding storage requirement without $ O $ is labeled near the running time.}\label{nqrand}}
\end{figure}
We randomly generate input quantum states as \cite{MISZCZAK2012118,qetlab} and $ M= d(d+1)  $.
With given number of copies $ N=3\times 10^4 $ for each output state and randomly generated process as in \eqref{4rand}, a comparison of results is shown in Fig. \ref{nqrand} where the horizontal axis is the number of qubits $ N_q $, the left vertical axis is the MSE on  a logarithmic scale and the right vertical axis is the running time $T$ (in seconds) on a logarithmic scale. We repeat the simulation $ 10 $ times and obtain the average MSE, the average running time and error bars.

When the two algorithms have a similar MSE (we set the error of convex optimization a little larger than TSS by changing CVX precision), the running time of TSS algorithm is smaller than convex optimization by approximately three orders of magnitude for three-qubit systems, indicating that our TSS algorithm is more efficient than the convex optimization method.
Since $ M=d(d+1) $ and we choose Cube bases where $ L=6^{N_q} $, the computational complexity for our algorithm is $ O(96^{N_q}) $.
For the simulated running time of
our algorithm, since our computational complexity analysis is asymptotical on $ d $, we fit the rightmost three
points and the slope  is $ 2.0843 $, which is close to the theoretical value $ \log _{10}(96)= 1.9823 $. The
slight difference  might be attributed to fact that the qubit number is
not large enough.
The storage requirement is also labeled by the black numbers in Fig. \ref{nqrand}. The storage requirement for our algorithm is $ O(ML) $. Since we use Cube bases of measurement, $ L=6^{N_q}, M=4^{N_q}+2^{N_q} $ and thus $ ML= 24^{N_q}+12^{N_q}$. As $ N_q $ increases from $ 1 $ to $ 5 $, $ ML $ are $36$, $720$, $15552$, $352512$ and $8211456$, respectively. For the convex  optimization method, the main storage requirement is the cost of storing $\mathcal O_{ml} $, which is a $ d^2\times d^2 $ matrix and the total number of these matrices is $ ML $.
Thus, the storage requirement is $ O(MLd^4) $.
As $ N_q $ increases from $ 1 $ to $ 3 $, $ MLd^{4}=384^{N_q}+192^{Nq} $ are $576$, $184320$ and $63700992$, respectively. Thus, our algorithm needs less storage than the convex optimization method.

\begin{figure}
	\centering
	\includegraphics[width=4.5in]{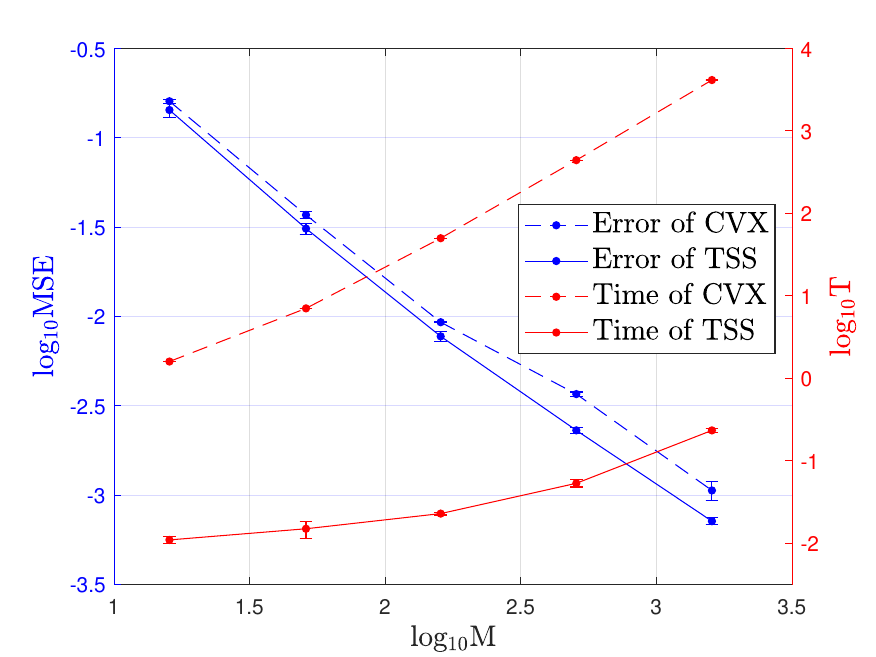}
	\centering{\caption{Using random quantum states, the running time $T$ and the MSE versus the different type of input states $M$ for the convex optimization method in \cite{unitary} and our TSS algorithm.}\label{typestate}}
\end{figure}
We then compare the two algorithms by changing the types of input states $ M $ with  $ d=4 $, where we
randomly generate input states as in \cite{MISZCZAK2012118,qetlab}  and the number of copies for each output state is $ N=3^{2}\times 10^{4} $. The comparison results are shown in Fig. \ref{typestate} where the horizontal axis is the different types of input states $ M $, the left vertical axis is the MSE  and the right vertical axis is the running time $T$, all on  a logarithmic scale.
From Fig. \ref{typestate}, the scaling between MSE and $ M $ is $-1.1474$, which is close to $ -1 $ and consistent with \eqref{statem}. This scaling is different from the scaling on $ N_t $ in Fig. \ref{mse} because here $ N $ is given and $ M $ increases, while in Fig. \ref{mse}, $ M $ is given and $ N $ increases.
For the simulated running time of
our algorithm,  we fit the rightmost three
points to obtain the slope $ 1.0090  $, which is close to the theoretical value $ 1 $. 
For the convex optimization method, the slope of the fitted line of the rightmost three points is $ 1.9186$, which is larger than that of our TSS algorithm, indicating that our TSS algorithm is more efficient than the convex optimization method.

\section{Conclusion}\label{sec8}

In this paper, we have introduced an analytical two-stage solution (TSS) applicable to both trace-preserving and non-trace-preserving QPT. Leveraging the natural basis, we have utilized the tensor structure of the coefficient matrix and provided insights into the computational complexity and storage requirements for our algorithm. Our contributions include the establishment of an analytical upper bound for errors and the optimization of input quantum states and measurement operators based on this error upper bound and condition number. The effectiveness and theoretical soundness of our algorithm have been demonstrated through numerical examples and testing on an IBM Quantum device.
 Furthermore, we have benchmarked our TSS algorithm against the convex optimization method outlined in \cite{unitary}, and the results show that our algorithm is more efficient in terms of both time and space costs.
Further work will focus on extending our algorithm to AAPT.

\section*{Acknowledgments}
The authors would like to thank Prof. Lloyd C. L. Hollenberg for the helpful discussion. This research was supported by the University of Melbourne through the establishment of the IBM Quantum Network Hub at the University. Shuixin Xiao would like to thank the support from the IEEE Control Systems Society Graduate Collaboration Fellowship.

\appendices
\section{Several propositions and lemmas}\label{appendixa}
Here, we  give some propositions and lemmas which are utilized in proving Theorem \ref{mainthe}.
\begin{proposition}\label{proposition2}
	For any $ A \in \mathbb C^{n\times k}  $ and $ b \in \mathbb C^{mk} $, we have
	\begin{equation}
	\left\|\left(I_{m} \otimes A\right) b\right\| \leq\|A\|\|b\|
	\end{equation}
\end{proposition}
\begin{IEEEproof}
	Let 
	\begin{equation}
	b=\left[b_{1}^{T}, b_{2}^{T}, \cdots, b_{m}^{T}\right]^{T},
	\end{equation}
	where $ b_i $ is a $ k\times 1 $ vector. Therefore,
	\begin{equation}
	\begin{aligned}
	\left\|\!\left(I_{m} \otimes A\right)\! b\right\|^2\!=\!\sum_{i=1}^{m}\left\|A b_{i}\right\|^2 
	\!\leq\!\|A\|^2 \sum_{i=1}^{m}\left\|b_{i}\right\|^2\!=\!\|A\|^2\|b\|^2.
	\end{aligned}
	\end{equation}
\end{IEEEproof}
	\begin{lemma}\cite{dbound}\label{lemma2}
	Let $\mathbb{H}_A$ and $\mathbb{H}_B$ be finite-dimensional Hilbert spaces of dimensions $d_{A}$ and $d_{B}$, respectively, and let $X \in \mathbb{H}_A \otimes \mathbb{H}_B$. Then for any unitarily invariant norm that is multiplicative over tensor products, the partial trace satisfies the norm inequality
\begin{equation}
\left\|\operatorname{Tr}_{A}(X) \right\| \leq \frac{d_{A}}{\left\|I_{A}\right\|}\|X\|,
\end{equation}
where $I_{A}$ is the identity operator.
\end{lemma}
\begin{lemma}(\cite{bhatia2007perturbation} Theorem 8.1 and Theorem 28.3)\label{lemma3}
	Let $X$, $Y$ be Hermitian matrices with eigenvalues $\lambda_1(X)\geq\cdots\geq\lambda_n(X)$ and $\lambda_1(Y)\geq\cdots\geq\lambda_n(Y)$, respectively. Then
	\begin{equation}\label{weyl}
	\max_j|\lambda_j(X)-\lambda_j(Y)|\leq||X-Y||,
	\end{equation}
	and
	\begin{equation}\label{weyl2}
	\sum_{j=1}^{n}\left(\lambda_j(X)-\lambda_j(Y)\right)^2\leq||X-Y||^2.
	\end{equation}
\end{lemma}
\section{SIC states and MUB states}\label{appendixb}
For  input states achieving the lower bounds of $M\operatorname{Tr}\big((V^{*} V^{T})^{-1}\big)$ and $ \operatorname{cond}(V) $, we give two examples, SIC states and MUB states, as the following propositions.
\begin{proposition}
	\label{lem1}
	$ d $-dimensional 	SIC  states (when they exist) belong to $ \underline{\mathrm{OIS}}(d,M) $ with the smallest $ M $ as $ M=d^2 $.	
\end{proposition}
\begin{proposition}
	\label{lem2}
	$ d $-dimensional MUB states (when they exist) belong to $ \underline{\mathrm{OIS}}(d,M) $ for $ M=d(d+1) $.	
\end{proposition}
The proofs are similar to the optimal probe states in quantum detector tomography and can be found in \cite{xiao2021optimal}.

For SIC-POVM in $ d=4 $, Bengtsson \cite{Bengtsson2010} gave one expression ignoring overall phases and normalization in the natural basis  as
	\begin{equation}
\label{sic}
\setlength{\arraycolsep}{1pt}
\left[\begin{array}{cccccccccccccccc}
x & x & x & x & \mathrm i &\mathrm  i & -\mathrm i & -\mathrm i & \mathrm i & \mathrm i & -\mathrm i & -\mathrm i & \mathrm i &\mathrm  i & -\mathrm i & -\mathrm i \\
1 & 1 & -1 & -1 & x & x & x & x & \mathrm i & -\mathrm i & \mathrm i & -\mathrm i & 1 & -1 & 1 & -1 \\
1 & -1 & 1 & -1 & 1 & -1 & 1 & -1 & x & x & x & x & -\mathrm i &\mathrm  i &\mathrm  i & -\mathrm i \\
1 & -1 & -1 & 1 & -\mathrm i &\mathrm  i &\mathrm  i & -\mathrm i & -1 & 1 & 1 & -1 & x & x & x & x
\end{array}\right],
\end{equation}	
where $ x=\sqrt{2+\sqrt{5}} $. Let $\left|\psi_{n}^{(\text{SIC})}\right\rangle$ be the $ n $-th column of \eqref{sic}. In this paper, we call the set of $$ \rho_n=\frac{\left|\psi_{n}^{(\text{SIC})}\right\rangle\left\langle\psi_{n}^{(\text{SIC})}\right|}{\operatorname{Tr}\left(\left|\psi_{n}^{(\text{SIC})}\right\rangle\left\langle\psi_{n}^{(\text{SIC})}\right|\right)} $$ as SIC states and $ \left\{P_n=\frac{1}{d}\rho_n \right\}_{n=1}^{d^2} $ as SIC-POVM.

For $ d=4 $, five MUB measurement sets  are
\begin{small}
	\begin{equation}
	\left\{\!|\psi_{n}^{(\text{MUB})}\rangle\!\right\}\!=\!\left\{\!\left\{\left|\psi_{n}^{A}\right\rangle\!\right\}\!,\!\left\{\left|\psi_{n}^{B}\right\rangle\!\right\}\!,\!\left\{\left|\psi_{n}^{C}\right\rangle\!\right\}\!,\!\left\{\left|\psi_{n}^{D}\right\rangle\!\right\}\!,\!\left\{\left|\psi_{n}^{E}\right\rangle\!\right\}\!\right\},
	\end{equation}
\end{small}
and
\begin{align*}
&\left\{\left|\psi_{n}^{A}\right\rangle\right\}=\{|00\rangle,|01\rangle,|10\rangle,|11\rangle\},\\
&\left\{\left|\psi_{n}^{B}\right\rangle\right\}=\{|R\pm\rangle,|L\pm\rangle\},\\
&\left\{\left|\psi_{n}^{C}\right\rangle\right\}=\{|\pm R\rangle,|\pm L\rangle\},\\
&\left\{\left|\psi_{n}^{D}\right\rangle\right\}=\left\{\frac{1}{\sqrt{2}}(|R 0\rangle \pm \mathrm i|L 1\rangle), \frac{1}{\sqrt{2}}(|R 1\rangle \pm \mathrm i|L 0\rangle)\right\},\\
&\left\{\left|\psi_{n}^{E}\right\rangle\right\}=\left\{\frac{1}{\sqrt{2}}(|R R\rangle \pm \mathrm i|L L\rangle), \frac{1}{\sqrt{2}}(|R L\rangle \pm \mathrm i|L R\rangle)\right\}, \numberthis \label{mubbase}
\end{align*}
where $|\pm\rangle=(|0\rangle \pm|1\rangle) / \sqrt{2}$, $|R\rangle=(|0\rangle-{\mathrm i}|1\rangle) / \sqrt{2}$, and
$|L\rangle=(|0\rangle+{\mathrm i}|1\rangle) / \sqrt{2}$ in the natural basis.
Adamson \emph{et al.}  \cite{PhysRevLett.105.030406} utilized these Mutually Unbiased Bases
in the QST. In this paper, we call $ \rho_n=|\psi_{n}^{(\text{MUB})}\rangle\langle\psi_{n}^{(\text{MUB})}| $  a MUB state and $ P_n=|\psi_{n}^{(\text{MUB})}\rangle\langle\psi_{n}^{(\text{MUB})}| $  a MUB measurement operator.

\bibliographystyle{ieeetr}         
\bibliography{process}

\end{document}